\documentclass[%
reprint,
amsmath,amssymb,
prb,
floatfix,
]{revtex4-2}

\usepackage{graphicx}
\usepackage{dcolumn}
\usepackage{bm}
\usepackage{bbold}
\usepackage{color}

\begin{document}

\title{Strictly one dimensional behavior  emerging from dispersive two dimensional system: implications on metallic nanowires on semiconducting substrates}
\author{Keshab Sony}
\affiliation{Leibniz Universit\"{a}t Hannover, Institute of Theoretical Physics,  Appelstr.~2, 30167 Hannover, Germany}
\author{Anas Abdelwahab}
\affiliation{Leibniz Universit\"{a}t Hannover, Institute of Theoretical Physics,  Appelstr.~2, 30167 Hannover, Germany}

\date{\today}

\begin{abstract}
We studied single and two Su-Shreiffer-Heeger wires on a simple cubic semiconducting substrate. The wire-wire coupling is either perpendicular or diagonal hopping respecting the particle-hole and time-reversal symmetries. The hybridization to the substrate renormalizes the model parameters of the wires towards the hopping parameter of the substrate without changing the basic nature of perpendicular or diagonal coupling and it can mediate effective perpendicular hopping but not diagonal hopping in the absence of direct wire-wire coupling. This justifies the investigation of multi uniform tight binding wires with perpendicular or diagonal hopping parameters while neglecting the substrate. Perpendicularly coupled uniform wires reveal an anisotropic two dimensional band dispersion. Diagonally coupled uniform wires reveal strictly one dimensional bands parallel to the wires direction if the intra-wire hopping parameter is larger than twice the diagonal hopping parameter despite strong dispersion perpendicular to the wires. Otherwise, they reveal strictly one dimensional bands parallel and perpendicular to the wires direction simultaneously. We established the possibility of realizing strictly one dimensional properties emerging from dispersive two dimensional systems if time-reversal and particle-hole symmetries are respected. This can facilitate the investigations of Luttinger liquid in the Au/Ge(001) and Bi/InSb(001) surface reconstructions.
\end{abstract}

\maketitle

\section{Introduction}\label{sec:Introduction}

Surface reconstructions of metallic atomic wires deposited on semiconducting substrates attracted lots of attention as platforms to realize features related to one-dimensional (1D) electrons, eg. Luttinger liquid phases~\cite{blu11,oht15}, Peierls metal-insulator transitions and charge-density-wave states~\cite{yeo99,aul13,che15}. For instance, the surface reconstruction Au/Ge(001) has been investigated and debated as a candidate to realize 1D correlated conductors for more than two decades~\cite{wan04,sch08,van08,van09,nak09,mey11,dud17}.
No consensus has been reached on the theoretically calculated exact structure of its surface reconstruction that match the experimental results~\cite{sau10,sei16,sei18}. One of the main issues is the apparent contradiction between the Luttinger liquid behavior reported experimentally, using scanning tunneling microscopy (STM) and spectroscopy (STS)~\cite{blu11,dud17}, and the strong dispersion perpendicular to the wires direction found in angle resolved photoemission spectroscopy (ARPES), STM and STS~\cite{nak11,par14,jon16,dud17}. Another system that reveals Luttinger liquid behavior is Bi deposited on InSb(001) surfaces in angle-resolved photoelectron spectroscopy~\cite{oht15}. However, this behavior is observed for large coverage of Bi on the InSb substrate, rendering the role of wire-wire coupling unclear.

Theoretically, the 1D correlated electrons are described primarily using "freestanding" 1D models which are then extended to anisotropic two- (2D) and three-dimensional (3D) systems~\cite{giamarchi04,schoenhammer04,gruener2000,solyom10}. However, metallic atomic wires on semiconducting substrates represent arrays of 1D wires coupled to a 3D reservoir, giving them a strong asymmetric nature and ruling out applicability of 2D and 3D anisotropic extensions. Therefore, it is necessary to investigate the influence of the coupling to the 3D bulk semiconducting substrate on the 1D features, and the possibility of substrate-mediated coupling between the wires.
A way of modelling was introduced~\cite{abd17a,abd17b,abd18,abd21} for single and two uniform atomic nanowires on a semiconducting substrate, amenable to investigations using methods for correlated electronic nanowires. However, the role of the substrate in mediating wire-wire coupling was not settled.

In this work, we use another approach to understand the impact of hybridization to the semiconducting substrate on the wires. This approach consists of two steps. The first step is to consider a single and two wires as topological insulators in the BDI class, namely as Su-Shreiffer-Heeger (SSH) wires. The two wires are coupled either with perpendicular or diagonal hopping, but not with both, in order to respect the symmetries of the BDI class. The substrate is described as a simple cubic lattice with conduction and valence orbitals at each site, such that it respects the required symmetries of the BDI class. We found that the hybridization to the substrate does not change the basic nature of the wires model parameters. However, it can effectively mediate perpendicular hopping but not diagonal hopping, in the absence of direct wire-wire coupling. These findings justify the extension of the number of wires while neglecting the substrate. Therefore, in the second step we analyze uniform multi wire systems coupled with either perpendicular or diagonal hopping, but not with both. We realized that the perpendicularly coupled wires can have the 1D properties as in usual anisotropic 2D conductors~\cite{giamarchi04,schoenhammer04,gruener2000,solyom10}. However, the diagonally coupled wires reveal strictly one dimensional bands, despite a strong dispersion in the perpendicular direction if they are extended in 2D. This finding demonstrates a way of possibility to realize strictly 1D behavior emerging from dispersive two dimensional uniform multi wire systems.

The paper is structured as follows. We introduce the model and the BDI class of topological insulators in sections~\ref{sec:model} and~\ref{sec:BDIclass}, respectively. In section~\ref{sec:SSHWiresNoSubstrate} we discuss free standing single, perpendicularly coupled and diagonally coupled SSH wires. In section~\ref{sec:SSHWiresWithSubstrate} we discuss the wires presented in section~\ref{sec:SSHWiresNoSubstrate} connected to simple cubic semiconducting substrate. We discuss uniform multi wire systems without a substrate in section~\ref{sec:MultiWireSystems}. Finally we conclude in section~\ref{sec:conclusion}.

\section{The model}\label{sec:model}
\begin{figure}[t]
\includegraphics[width=0.4\textwidth]{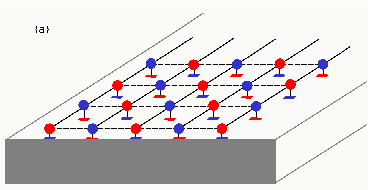}
\includegraphics[width=0.4\textwidth]{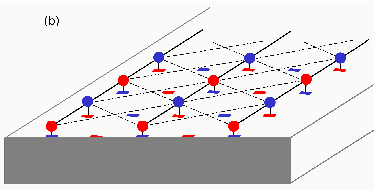}
\caption{\label{fig:WiresOnSubstrate} (a) Nearest neighbor wires on substrate that can be perpendicularly coupled (b) Next nearest neighbor wires on substrate that can be diagonally coupled.
}
\end{figure}

We consider wire-substrate systems which are transnational invariant along the wires direction, described by the Hamiltonian
\begin{equation}\label{eq:FullHamiltonian}
 H = H_{\text{sbt}} + H_{\text{wires}} + H_{\text{hyb}} ,
\end{equation}
where $H_{\text{sbt}}$ describes the substrate, $H_{\text{wires}}$ describes the wires and $H_{\text{hyb}}$ describes the hybridization between the wires and the substrate.
We label the repeated unit cell along the wires direction by $u$, such that the total number of unit cells is $N_u$. We restrict the coordinates $(x,y,z)$ to be inside the unit cell $u$. The thermodynamic limit corresponds to $N_u \rightarrow \infty$. The periodic boundary conditions (PBC) along the wires direction ($x$-direction) corresponds to finite $N_u$, such that $u=1=1+N_u$. The open boundary conditions (OBC) along the same direction corresponds to finite $N_u$, such that $u=N_u+1=0$.
We consider arbitrary number of SSH wires coupled to a semiconducting substrate as shown in Fig.~\ref{fig:WiresOnSubstrate}, where the number of wires is denoted $N_{\text{w}}$.  

The substrate Hamiltonian is given by
\begin{equation}\label{eq:substrate}
 H_{\text{sbt}} = \sum_{\text{s}=\text{v},\text{c}} H_{\text{s}} ,
\end{equation}
where the subscript v represent the orbitals of the valence bands, and c represent the orbitals of the conduction bands, such that
\begin{eqnarray}\label{eq:substrate_s}
H_{\text{s}} &=& \sum_{u,\mathbf{r}} \epsilon_{\text{s}} n_{u,\text{s}, \mathbf{r} } \nonumber \\
&+& t_{\text{s}} \sum_{u,\langle \mathbf{r} \mathbf{q} \rangle}  \left (
c^{\dag}_{u,\text{s}, \mathbf{r}}  c^{\phantom{\dag}}_{u,\text{s},\mathbf{q}} + \text{H.c.}
\right ) \nonumber \\
&+& t_{\text{s}} \sum_{u,\langle \mathbf{r} \mathbf{q} \rangle}  \left (
c^{\dag}_{u,\text{s}, \mathbf{q}}  c^{\phantom{\dag}}_{u+1,\text{s},\mathbf{r}} + \text{H.c.}
\right ) .
\end{eqnarray}
The operator $c^{\dag}_{u,\text{s},\mathbf{r}}$ creates a spinless fermion on the orbital $\text{s}$ localized on the site with coordinates $\mathbf{r} = (x,y,z)$, such that $x=1,...,\frac{L_x}{N_u}$, $y=1,...,L_y$ and $z=1,...,L_z$, where $L_x$, $L_y$ and $L_z$ are total lengths in $x$- $y$- and $z$- directions, respectively. The substrate fulfills the PBC in the $y$ direction, ie. $y=1=1+L_y$, and the OBC in the $z$ direction, ie. it terminates at $z=L_z$.
The fermion density operator on each orbital is $n_{u,\text{s} ,\mathbf{r}} = c^{\dag}_{u,\text{s}, \mathbf{r}} c^{\phantom{\dag}}_{u,\text{s}, \mathbf{r}}$.
The first term in Eq.(\ref{eq:substrate_s}) set the local potential $\epsilon_{\text{s}}=\epsilon_{\text{v}}$ for the orbitals of valence bands, and $\epsilon_{\text{s}}=\epsilon_{\text{c}}$ for the orbitals of conduction bands. The second term set the intra unit cell hopping and the third term set the inter unit cell hopping. We set $t_{\text{c}}=t_{\text{v}}=t_{\text{s}}$ and chose $t_{\text{s}}$ as energy unit. We set $\epsilon_{\text{c}}=-\epsilon_{\text{v}}=7$, such that a band gap is open in the substrate band structure.
Despite the homogeneous hopping terms in the conduction and the valence bands of the substrate, we distinguish between the intra and inter unit cell hopping terms in the substrate due to the dimerization in the wires. Therefore, we get $\frac{L_x}{N_u}=2$.

The Hamiltonian of the wires is given by
\begin{equation}\label{eq:WiresHamiltonian}
 H_{\text{wires}} = \sum_{\text{w}=1,...,N_{\text{w}}} H_{\text{w}} + \sum_{\text{w}=1,...,N_{\text{w}-1}} H_{\text{w},\text{w}+1},
\end{equation}
where $H_{\text{w}}$ represents a SSH wire given by
\begin{eqnarray}\label{eq:HamiltonianSSH}
H_{\text{w}} &=&  \sum_{u} t_{\text{w}} \left( c^{\dag}_{u,r_{\text{w}}} c^{\phantom{{\dag}}}_{u,q_{\text{w}}} + \text{H.c.} \right) 
\nonumber \\
&+& \sum_{u} t^{\prime}_{\text{w}} \left( c^{\dag}_{u,q_{\text{w}}} c^{\phantom{{\dag}}}_{u+1,r_{\text{w}}} + \text{H.c.} \right) .
\end{eqnarray}
Here $c^{\dag}_{u,r_{\text{w}}}$ and $c^{\dag}_{u,q_{\text{w}}}$ $\left( c^{\phantom{{\dag}}}_{u,r_{\text{w}}} \text{ and } c^{\phantom{{\dag}}}_{u,q_{\text{w}}}\right)$ denote the creation (annihilation) operators for a spinless fermion in unit cell $u$,
where $r_{\text{w}}=(1,y_{\text{w}},0)$ and $q_{\text{w}}=(2,y_{\text{w}},0)$.
We set $t_{\text{w}}=t_{\text{s}}+\delta_{\text{w}}$ and $t_{\text{w}}^{\prime}=t_{\text{s}}-\delta_{\text{w}}$, where the dimerization is given by setting $-t_{\text{s}} \leq \delta_{\text{w}}\leq t_{\text{s}}$. For simplicity, we set $\delta_{\text{w}}=\delta$ for all w, ie. $t_{\text{w}}=t$ and $t_{\text{w}}^{\prime}=t^{\prime}$.
Needless to mention, that for a single wire system we omit the summation over $\text{w}$.

In the case of a multi wire system, there can be perpendicular or diagonal coupling between adjacent wires. The perpendicular coupling is given by
\begin{eqnarray}\label{eq:perpendicular-hopping}
 H_{\perp} &=& \sum_{\text{w} = 1,...,N_{\text{w}-1}} H_{\text{w},\text{w}+1} \nonumber \\
 &=&  \sum_{u} t_{\perp} \left( c^{\dag}_{u,r_{\text{w}}} c^{\phantom{{\dag}}}_{u,r_{\text{w}+1}} + \text{H.c.} \right)
\nonumber \\
&+&  \sum_{u} t_{\perp} \left( c^{\dag}_{u,q_{\text{w}}} c^{\phantom{{\dag}}}_{u,q_{\text{w}+1}} + \text{H.c.} \right) ,
\end{eqnarray}
and the diagonal coupling is given by
\begin{eqnarray}\label{eq:diagonal-hopping}
 H_d &=& \sum_{\text{w}=1,...,N_{\text{w}-1}} H_{\text{w},\text{w}+1} \nonumber \\
 &=&  \sum_{u} t_d \left( c^{\dag}_{u,r_{\text{w}}} c^{\phantom{{\dag}}}_{u,q_{\text{w}+1}} + \text{H.c.} \right)
\nonumber \\
&+&  \sum_{u} t_d \left( c^{\dag}_{u,q_{\text{w}}} c^{\phantom{{\dag}}}_{u,r_{\text{w}+1}} + \text{H.c.} \right)
\nonumber \\
&+&  \sum_{u} t_d \left( c^{\dag}_{u,q_{\text{w}}} c^{\phantom{{\dag}}}_{u+1,r_{\text{w}+1}} + \text{H.c.} \right)
\nonumber \\
&+& \sum_{u} t_d \left( c^{\dag}_{u,q_{\text{w}+1}} c^{\phantom{{\dag}}}_{u+1,r_{\text{w}}} + \text{H.c.} \right) . 
\end{eqnarray}
It is important to mention that adjacent wires labeled by $\text{w}$ and $\text{w}+1$ are not necessarily hybridized to nearest neighbors sites on the surface of the substrate, ie. $y_{\text{w}+1}-y_{\text{w}}\geq 1$.

The hybridization between each wire and the substrate is given by
\begin{equation}\label{eq:Hybridization}
 H_{\text{hyb}} = \sum_{\text{w,s}} H_{\text{w,s}} ,
\end{equation}
where
\begin{eqnarray}\label{eq:Hybridization2}
H_{\text{w,s}} &=& \sum_{u} t_{\text{w,s}} \left( c^{\dag}_{u,{\text{s}}, \mathbf{r}}
c^{\phantom{\dag}}_{u,r_{\text{w}}} + \text{H.c.}  \right) \nonumber \\ 
&+& \sum_{u} t_{\text{w,s}} \left( c^{\dag}_{u,{\text{s}}, \mathbf{q}}
c^{\phantom{\dag}}_{u,q_{\text{w}}} + \text{H.c.}  \right)
\end{eqnarray}
represents the hybridization of the wire sites $r_{\text{w}} = (1,y_{\text{w}},0)$ and $q_{\text{w}} = (2,y_{\text{w}},0)$, with the orbitals of valence and conduction bands on sites $\mathbf{r} = (1,y_{\text{w}},1)$ and $\mathbf{q} = (2,y_{\text{w}},1)$ of the substrate. We set $t_{\text{w,v}}=t_{\text{w,c}}=t_{\text{ws}}$.

For noninteracting wire-substrate systems with PBC, $H$ can be written as a sum of commuting operators $H(k_j)$, acting only on the single-particle Bloch states, with the wave number 
\begin{equation}\label{eq:wavenumber}
k_j = \frac{2\pi j}{N_u}
\end{equation}
in the first Brillouin zone, where the quantum number $j$ satisfies $ -N_u/2 \leq j < N_u/2$.
The transformation of the full wire-substrate Hamiltonian~(\ref{eq:FullHamiltonian}) to the momentum space in $x$-direction is performed using the canonical transformation
\begin{equation}\label{eq:TransformationMomentumSpace}
 c^{\phantom{{\dag}}}_{u,\text{R}} = \frac{1}{\sqrt{N_u}} \sum_{j} c^{\phantom{{\dag}}}_{k_j,\text{R}} \exp\left(-ik_j u\right) ,
\end{equation}
where $\text{R}=(x,y,z)$ represents the coordinates inside the unit cell $u$. Unless it is explicitly stated, we omit the quantum number $j$. Therefore, the substrate part of the Hamiltonian is given by $H_{\text{sbt}}=\sum_k H_{\text{sbt}}(k)$, such that
\begin{eqnarray}
H_{\text{sbt}}(k) &=& \sum_{\text{s}=\text{v},\text{c}} H_{\text{s}}(k) , \nonumber \\
 H_{\text{s}}(k) &=&  \epsilon_{\text{s}} \sum_{\mathbf{r}} n_{k,\text{s}, \mathbf{r} } \nonumber \\
&+&   t_{\text{s}} \sum_{\langle \mathbf{r} \mathbf{q} \rangle}  \left (
c^{\dag}_{k,\text{s}, \mathbf{r}}  c^{\phantom{\dag}}_{k,\text{s},\mathbf{q}} + \text{H.c.}
\right )   \nonumber \\
&+&  t_{\text{s}} \sum_{\langle \mathbf{r} \mathbf{q} \rangle}  \left (
c^{\dag}_{k,\text{s}, \mathbf{q}}  c^{\phantom{\dag}}_{k,\text{s},\mathbf{r}} \exp(-ik) + \text{H.c.}
\right ) .  
\end{eqnarray}
The wire-substrate hybridization part is given by $H_{\text{hyb}}=\sum_k H_{\text{hyb}}(k)$,
where
\begin{eqnarray}
\label{eq:HybridizationMomentum}
H_{\text{hyb}}(k) &=& \sum_{\text{w}=1,...,N_{\text{w}}} \sum_{\text{s}=\text{v},\text{c}} H_{\text{ws}}(k) , \nonumber \\
H_{\text{ws}}(k) &=& t_{\text{ws}} \left( c^{\dag}_{k,{\text{s}}, \mathbf{r}}  
c^{\phantom{\dag}}_{k,r_{\text{w}}} + \text{H.c.}  \right) \nonumber \\
&+& t_{\text{ws}} \left( c^{\dag}_{k,{\text{s}}, \mathbf{q}}  
c^{\phantom{\dag}}_{k,q_{\text{w}}} + \text{H.c.}  \right) .
\end{eqnarray}

As we stated before, we investigate how the wire-substrate hybridization affect the intra wire hopping parameters, and whether it changes, preserves or mediates wire-wire couplings. Therefore, for the sake of simplicity, we will discuss only single and two adjacent wires coupled either by perpendicular or diagonal hopping. The transformation of the single and two wires to the momentum space will be discussed in section~\ref{sec:SSHWiresNoSubstrate}.

\section{Topological insulators in the BDI class}\label{sec:BDIclass}

To investigate the impact of the wire-substrate hybridization on the properties of the wires, we consider the full wire-substrate system as a topological insulator~\cite{shen12,asb16,ber18}. This allows us to detect changes reflected on the topological properties of the system. Band insulators are topologically classified according to the so called periodic table of the topological insulators~\cite{sch08ClassTI,sch09ClassTI,kitaevTI,ryu10}. This classification depends on the dimensionality, as well as whether one or more of the following three symmetries are fulfilled. 

The first symmetry is the chiral symmetry, which is present due to the bipartite nature of the lattice, therefore, it is also named sublattice symmetry. It guarantees the existence of Hermitian unitary operator $S$ acting within the unit cell and anticommuting with the Hamiltonian, such that, for the Bloch Hamiltonian $H(k)$,
\begin{equation}\label{eq:chiral}
    S H(k) S = -H(k) ,
\end{equation}
where $k\in \{k_j\}$ as defined in Eq.~(\ref{eq:wavenumber}). Thus, $H(k)$ can be written in a completely block off diagonal form
\begin{equation}
\label{eq:OffDiagonalForm}
 H(k) =  
\begin{bmatrix}
0& h(k)\\
h^{\dagger}(k)& 0
\end{bmatrix} .
\end{equation}
This symmetry restricts the arrangement of the wires on the surface of the cubic lattice substrate in the wire-substrate model~(\ref{eq:FullHamiltonian}). Specifically, it does not allow simultaneous perpendicular and diagonal coupling between the wires. In the perpendicularly (diagonally) coupled wires, the laterally adjacent wire sites can only be connected with substrate sites that belong to different (similar) sublattice, see Fig.~\ref{fig:WiresOnSubstrate}. The chiral symmetry can be present without requiring other symmetries. However, systems that fulfill simultaneously time reversal and particle hole symmetries must fulfill the chiral symmetry. These are the two remaining symmetries required for the BDI class.

For spinless fermions, time reversal symmetry is defined by the antiunitary operator $T$, which merely takes the complex conjugate, such that
\begin{equation}\label{eq:TimeReversal}
    T H(k) T^{-1} = H(-k) ,
\end{equation}
as long as the model parameters are restricted to real values. The particle-hole symmetry for spinless fermions is defined by the symmetry under the transformation $c^{\dag}_{u,\text{R}}=F_{u,\text{R}}c^{\phantom{{\dag}}}_{u,\text{R}}$, where $F_{u,\text{R}}=-1$ if the site on $\text{R}$ belongs to one sublattice, and $F_{u,\text{R}}=1$ if it belongs to the other sublattice. This lead to the antiunitary operator $P$ that satisfy
\begin{equation}\label{eq:ParticleHole}
    P H(k) P^{-1} = -H(-k) .
\end{equation}
The wire-substrate model fulfills the time reversal and particle-hole symmetries such that $T^2=P^2=\mathbb{1}$.
The topological insulators in BDI class can have nontrivial topological phases, if they have 1D but not 2D or 3D band structure.

Topological phases of the BDI class are characterized by a topological invariant called the winding number $W\in\mathbb{Z}$~\cite{shen12,asb16,ber18}. Crossing between phases with different $W$ is a topological phase transition, which is accompanied with closing the band gap rendering the system critical at the phase transition. $W$ can be defined, for systems with PBC, as the winding number of the graph $\text{Det}(h(k))$ around the origin of the complex plane, for $k \in \left[-\pi, \pi \right)$, where $h(k)$ is the block off diagonal matrix defined in Eq.~(\ref{eq:OffDiagonalForm}). It can be obtained using
\begin{equation}\label{eq:WindingNumber}
    W = \frac{1}{2i\pi} \int^{\pi}_{-\pi} \frac{\partial}{\partial k} \text{log}\left[\text{Det}(h(k))\right] dk .
\end{equation}
Trivial topological phases correspond to $W=0$, while nontrivial topological phases corresponds to $W\neq 0$. $|W|$ gives the number of edge localized states at energy $E=0$ for systems with OBC in the thermodynamic limit.
Sometimes, it is not trivial to transform $H(k)$ into the off-diagonal form in Eq.~(\ref{eq:OffDiagonalForm}). Nevertheless, a way to obtain $W$~\cite{van18} is by constructing the overlap matrix $U$ from the occupied energy eigenvectors $| u(k) \rangle$, such that 
\begin{equation}\label{eq:OverlapMatrix}
 U_{n,m}(k)=\langle u_n(k)|\frac{\partial}{\partial k}u_m(k) \rangle .
\end{equation}
By integrating over the Brillouin zone, we get
\begin{equation}\label{eq:WindingNumber2}
 W = \frac{1}{i\pi} \int^{\pi}_{-\pi} \text{Det} \left[ U_{n,m}(k) \right] dk .
\end{equation}

\section{Free standing SSH wires}\label{sec:SSHWiresNoSubstrate}
\subsection{Band structures and single particle spectral functions}\label{sec:BandsSpectralFunctionsNoSubstrate}
\begin{figure}[t]
\includegraphics[width=0.27\textwidth]{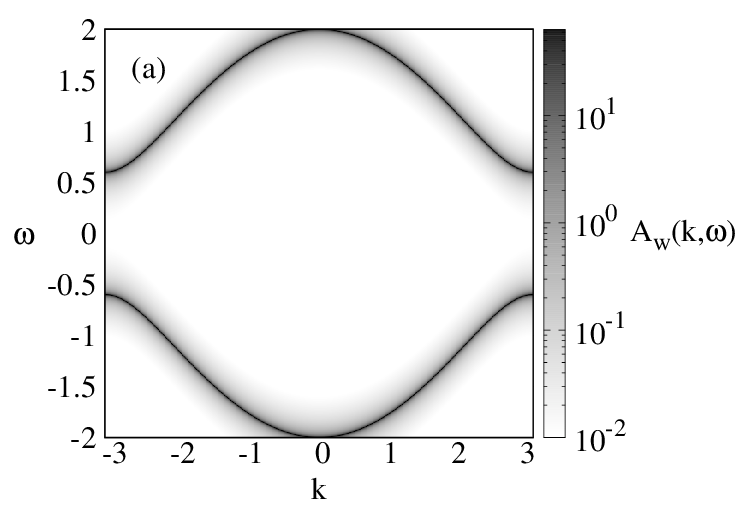}
\includegraphics[width=0.27\textwidth]{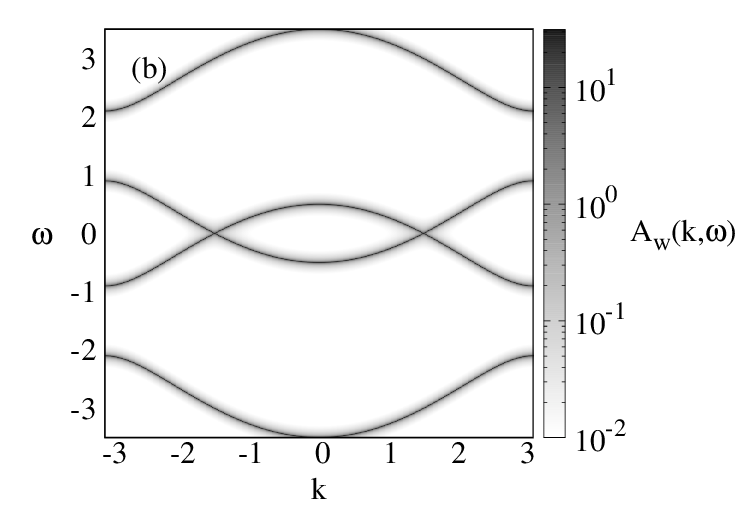}
\includegraphics[width=0.27\textwidth]{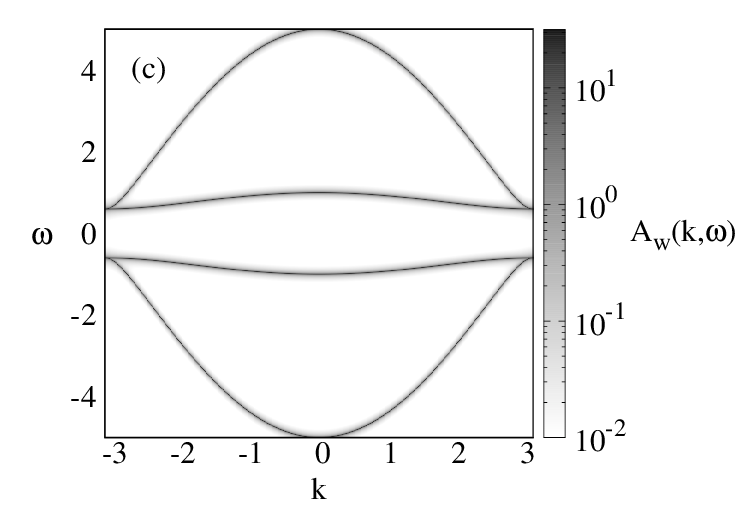}
\caption{\label{fig:SpectralFunctionsNoSubstrate} Single particle spectral function for single and two SSH wires with $N_u=500$ and $\delta=-0.3$. (a) single wire (b) two perpendicularly coupled wires with $t_{\perp}=1.5$ (c) two diagonally coupled wires with $t_d=1.5$.
}
\end{figure}

The free standing single SSH wire without a substrate is described by reducing Hamiltonian~(\ref{eq:FullHamiltonian}) to $H = H_{\text{w}}$ in~(\ref{eq:HamiltonianSSH}). It is a well known example of topological insulators~\cite{shen12,asb16,ber18}. By transforming the single wire Hamiltonian to momentum space, we get its matrix form
\begin{equation}
\label{eq:HkSingleChain}
 H(k) =
\begin{bmatrix}
0 & t+t^{\prime}e^{ik} \\
t+t^{\prime}e^{-ik} & 0
\end{bmatrix} ,
\end{equation}
such that $h(k)=t+t^{\prime}e^{ik}$.
We can obtain square of energy bands $E^2_{l}(k)$ by diagonalizing either
\begin{equation}\label{eq:hk_Hdagk}
 \hat{H}(k)=h^{\dag}(k)h(k)
 \hspace{5mm} \text{or} \hspace{5mm}
 \bar{H}(k)=h(k)h^{\dag}(k) ,
\end{equation}
where $l$ is the band index~\cite{maf18}.
Thus, for the single SSH wire, we obtain the two energy bands
\begin{equation}
 \label{eq:EkSingleChain}
 E_{l}(k) = \pm \sqrt{t^2 + t^{\prime 2} + 2tt^{\prime}\cos(k)} .
\end{equation}
The band gap is given by
\begin{equation}\label{eq:BandGapSingleSSHWire}
E_g = 4|\delta| ,
\end{equation}
while the top (bottom) of the upper (lower) band is fixed at $E_{upper}(0)=2$ ($E_{lower}(0)=-2$).
We can display the band structures using the single particle spectral function defined as
\begin{equation}\label{eq:Aw}
 A(k,\omega) = \sum_{l} \delta(\omega - E_{l}(k)) ,
\end{equation}
where $\delta(...)$ is Dirac delta function.
Such spectral function is equivalent to the spectral dispersions seen in the angle resolved photoemission spectroscopy. To draw $A(k,\omega)$, we substitute the Dirac delta function $\delta(...)$ by the Lorentzian function with $\eta=0.005$. Figure~\ref{fig:SpectralFunctionsNoSubstrate}(a) display the single particle spectral function for a free standing single SSH wire with $\delta=-0.3$. We clearly observe the band dispersions given by Eq.~(\ref{eq:EkSingleChain}).

The two perpendicularly coupled SSH wires without a substrate are described by $H = \sum_{\text{w}=1,2} H_{\text{w}}+H_{\perp}$.
They possess a mirror reflection symmetry between the two legs around the line crossing the rungs midpoints.
Therefore, it is more illustrative to transform the Hamiltonian to the basis of bonding, $d^{(-)\phantom{{\dag}}}_{u,x}$, and anti-bonding, $d^{(+)\phantom{{\dag}}}_{u,x}$, operators, defined as
\begin{eqnarray}
\label{eq:Bonding-AntiBondingOp}
    d^{(-)\phantom{{\dag}}}_{u,x} &=& \frac{1}{\sqrt{2}} \left( c^{\phantom{{\dag}}}_{u,r_1} - c^{\phantom{{\dag}}}_{u,r_2} \right) \nonumber \\
    d^{(+)\phantom{{\dag}}}_{u,x} &=& \frac{1}{\sqrt{2}} \left( c^{\phantom{{\dag}}}_{u,r_1} + c^{\phantom{{\dag}}}_{u,r_2} \right) .
    \end{eqnarray}
Therefore, we get 
\begin{eqnarray}\label{eq:HamiltonianPerpBondingAntibonding}
 H^{(\mp)} &=& \sum_{u} t \left( d^{(\mp)\dag}_{u,1} d^{(\mp)\phantom{{\dag}}}_{u,2} + \text{H.c.} \right)
\nonumber \\
&+& \sum_{u} t^{\prime} \left( d^{(\mp)\dag}_{u,2} d^{(\mp)\phantom{{\dag}}}_{u+1,1} + \text{H.c.} \right)  
\nonumber \\
&\mp& t_{\perp} \sum_{u} d^{(\mp)\dag}_{u,1}d^{(\mp)\phantom{{\dag}}}_{u,1} \mp t_{\perp} \sum_{u} d^{(\mp)\dag}_{u,2}d^{(\mp)\phantom{{\dag}}}_{u,2}  ,
\end{eqnarray}
where $H^{(-)}$ ($H^{(+)}$) acts on bonding (antibonding) orbitals.
By transforming $H^{(-)}$ and $H^{(+)}$ to momentum space and diagonalizing them, we obtain the four bands
\begin{eqnarray}
 \label{eq:EkTwoPerpLadderBonding}
 E^{(-)}_{l}(k) &=& -t_{\perp} \pm \sqrt{t^2 + t^{\prime 2} + 2tt^{\prime}\cos(k)} \nonumber \\ 
 E^{(+)}_{l}(k) &=& t_{\perp} \pm \sqrt{ t^2 + t^{\prime 2} + 2tt^{\prime}\cos(k)} ,
\end{eqnarray}
where $E^{(\mp)}_{l}(k)$ are the dispersions of bonding and antibonding bands, respectively.
Both bonding and antibonding bands are identical to the bands of the free standing single SSH wire in the absence of perpendicular hopping. The perpendicular hopping acts as chemical potential with opposite signs, shifting the bonding and antibonding bands, such that the top (bottom) of the upper (lower) bonding band is given by $E^{(-)}_{upper}(0)=2-t_{\perp}$ ($E^{(-)}_{lower}(0)=-2-t_{\perp}$), while the top (bottom) of the upper (lower) antibonding band is given by $E^{(+)}_{upper}(0)=2+t_{\perp}$ ($E^{(+)}_{lower}(0)=-2+t_{\perp}$). The inner band gap inside the bonding and antibondig bands is given by $E^{(\mp)}_g = 4|\delta|$, matching the band gap of a single SSH wire. The global band gap is given by
\begin{equation}\label{eq:BandGapTwoSSHWiresPerp}
    E_g = 
\begin{cases}
    4|\delta|-2t_{\perp},& \text{if } t_{\perp} < 2 \text{ and } t_{\perp} < 2|\delta|\\
    -4+2t_{\perp},& \text{if } t_{\perp} > 2 ,
\end{cases}
\end{equation}
otherwise, the system is gapless.
Each of $H^{(-)}$ and $H^{(+)}$, separately, breaks the particle-hole symmetry. In fact, the transformation to bonding and antiboding representation mixes the two sublattices in the perpendicularly coupled wires. However, the lower (upper) bonding band is the chiral partner of the upper (lower) antibondig band, rendering the full system particle-hole symmetric.
Figure~\ref{fig:SpectralFunctionsNoSubstrate}(b) shows the single particle spectral function of perpendicularly coupled two SSH wires with $\delta=-0.3$ and $t_{\perp}=1.5$. We clearly observe the energy shift by $-1.5$ ($1.5$) of the bonding (antibonding) bands in comparison with the single SSH wire bands.

The two diagonally coupled SSH wires without a substrate are described by $H = \sum_{\text{w}=1,2} H_{\text{w}}+H_d$.
Using the bonding and antibonding operators, we transform the Hamiltonian into the two bonding and antibonding Hamiltonians
\begin{eqnarray}\label{eq:HamiltonianDiagBondingAntibonding}
 H^{(\mp)} &=& \sum_{u} t_{(\mp)} \left( d^{(\mp)\dag}_{u,1} d^{(\mp)\phantom{{\dag}}}_{u,2} + \text{H.c.} \right)
\nonumber \\
&+& \sum_{u} t^{\prime}_{(\mp)} \left( d^{(\mp)\dag}_{u,2} d^{(\mp)\phantom{{\dag}}}_{u+1,1} + \text{H.c.} \right)  .
\end{eqnarray}
Here $t_{(-)}=t- t_d$ and $t^{\prime}_{(-)}=t^{\prime}-t_d$ for the bonding Hamiltonian, while $t_{(+)}=t+ t_d$ and $t^{\prime}_{(+)}=t^{\prime}+t_d$ for the antibonding Hamiltonian. Each of $H^{(-)}$ and $H^{(+)}$, separately, resembles an independent single SSH wire and respects all the required symmetries for the BDI class. The transformation to bonding and antiboding representation does not mix the two sublattices in the diagonally coupled wires. Therefore, the reflection symmetry imposes additional requirements to classify the topological phases, as we will see in Sec.~\ref{sec:PhaseDiagramNoSubstrate}.
By transforming Hamiltonians~(\ref{eq:HamiltonianDiagBondingAntibonding}) to the momentum space and diagonalizing them, we get the two bonding bands
\begin{equation}
 \label{eq:EkTwoDiagLadderBonding}
 E^{(-)}_{l}(k) = \pm \sqrt{t_{(-)}^2 + t^{\prime2}_{(-)} + 2t^{\phantom{{\prime}}}_{(-)} t^{\prime}_{(-)}\cos(k)} ,
 \end{equation}
 and the two antibonding bands
 \begin{equation}\label{eq:EkTwoDiagLadderAntibonding}
 E^{(+)}_{l}(k) = \pm \sqrt{t_{(+)}^2 + t^{\prime2}_{(+)} + 2t^{\phantom{{\prime}}}_{(+)} t^{\prime}_{(+)}\cos(k)} .
\end{equation}
The inner band gap inside the bonding and antibondig bands is equal to the global band gap given by Eq.~(\ref{eq:BandGapSingleSSHWire}), indicating that all the upper (and lower) bonding and antibonding bands matches at $E_{l}(\pm\pi)$.
The top (bottom) of the upper (lower) bonding band is given by $E^{(-)}_{upper}(0)=2t_{(-)}$ ($E^{(-)}_{lower}(0)=-2t_{(-)}$), while the top (bottom) of the upper (lower) antibonding band is given by $E^{(+)}_{upper}(0)=2t_{(+)}$ ($E^{(+)}_{lower}(0)=-2t_{(+)}$). The bonding bands become completely flat at $t_d=t$ or $t_d=t^{\prime}$.
Figure.~\ref{fig:SpectralFunctionsNoSubstrate}(c) display the single particle spectral function of two diagonally coupled wires with $t_d=1.5$ and $\delta=-0.3$. We observe clearly the energy dispersions given by Eq.~(\ref{eq:EkTwoDiagLadderBonding}) and~(\ref{eq:EkTwoDiagLadderAntibonding}).

\subsection{Phase diagrams}\label{sec:PhaseDiagramNoSubstrate}
The single SSH wire has a nontrivial (trivial) topological phase for $\delta>0$ ($\delta<0$), with $W=1$ ($W=0$) identified using Eq.~(\ref{eq:WindingNumber}), with critical point at $\delta=0$.
\begin{figure}[t]
\includegraphics[width=0.27\textwidth]{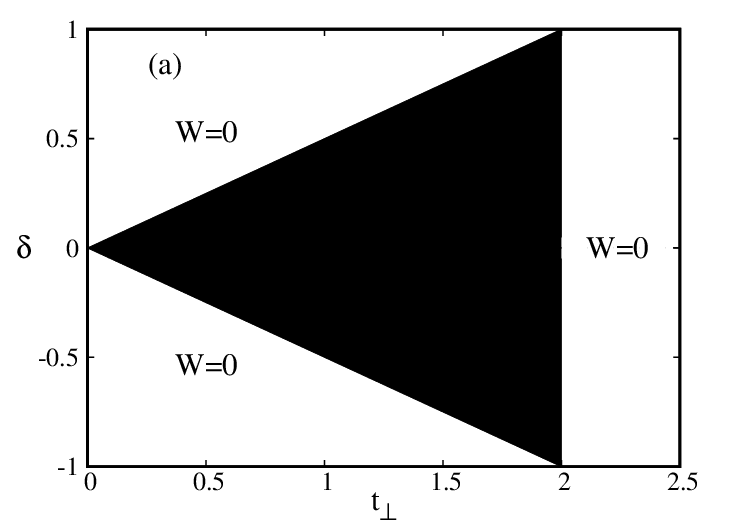}
\includegraphics[width=0.27\textwidth]{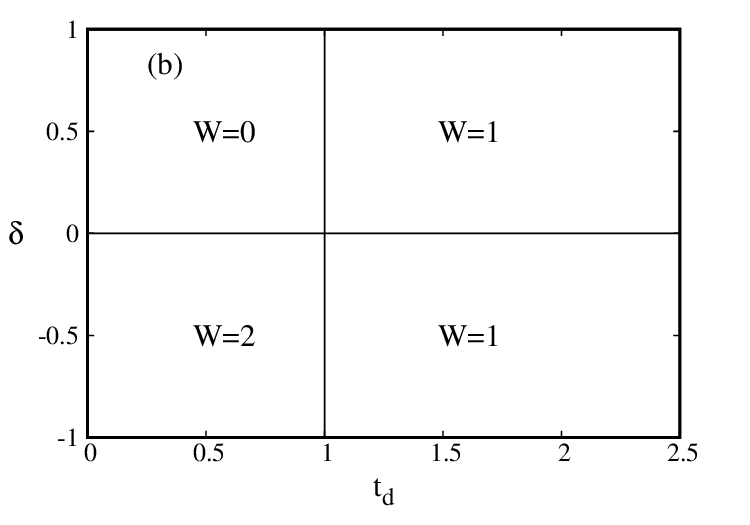}
\caption{\label{fig:PhaseDiagramNoSubstrate} Topological phase diagrams of (a) two perpendicularly coupled SSH wires and (b) two diagonally coupled SSH wires.
}
\end{figure}
As we realized before, the bonding and antibondig Hamiltonians of the perpendicularly coupled SSH wires break the particle-hole symmetry if they are considered separately. Therefore, we transform the full Hamiltonian of the two wires to momentum space, and apply the chiral transformation to get the off-diagonal part
\begin{equation}\label{eq:hk2LegPerpHopping}
 h(k) =
 \begin{bmatrix}
t+t^{\prime}e^{-ik} & t_{\perp} \\
 t_{\perp} & t+t^{\prime}e^{ik}
\end{bmatrix} .
\end{equation}
It is clear that, $\forall k \text{ Det}(h(k)) \in \mathbb{R}$, hence there is no topologically nontrivial phase for the perpendicularly coupled SSH wires. The phase diagram contains either regions with trivial topological phase or with gapless phase, given by Eq.~(\ref{eq:BandGapTwoSSHWiresPerp}) and shown in Fig.~\ref{fig:PhaseDiagramNoSubstrate}(a).
Due to the preservation of the symmetries required for the BDI class in each of the bonding and antibonding Hamiltonians of the diagonally coupled SSH wires,
we can define the winding number $W_-$ ($W_+$) of the bonding  (antibonding) bands. We realize that, for $t_d>t_{\text{s}}$, there is a gap closing at $\delta=0$ without changing the winding number $W$. This is due to the additional reflection symmetry, which imposes a second topological index, namely the difference $\Delta W=W_- - W_+$, that distinguishes between the topological nature of the bonding and antibondig bands.
Therefore, we can distinguish four topological phases in the phase diagram of the diagonally coupled wires.
In the first phase with $t_d<t_{\text{s}}$ and $\delta>0$, both bonding and antibonding bands are trivial, hence, $W=0$ and $\Delta W=0$. The second phase is for $t_d<t_{\text{s}}$ and $\delta<0$, where both bands are nontrivial, hence, $W=2$ and $\Delta W=0$. In the third phase with $t_d>t_{\text{s}}$ and $\delta>0$, the bonding band is nontrivial while the antibonding band is trivial, hence,  $W=1$ and $\Delta W=1$. In the forth phase with $t_d>t_{\text{s}}$ and $\delta<0$, the bonding band is trivial while the antibonding band is nontrivial, hence,  $W=1$ and $\Delta W=-1$.
The phase diagram of the diagonally coupled SSH wires is displayed in Fig.~\ref{fig:PhaseDiagramNoSubstrate}(b). We mention here that the diagonally coupled wires with $\delta=0$ and $t_d<t_{\text{s}}$ should reveal topological phases at criticality according to~\cite{ver18,ver20}. However, this is out the scope of our investigation, thus we postponed it to future investigation.

\subsection{Energy spectrum and local density of states for wires with OBC}\label{sec:EnergySpectrumLDOS-NoSubstrateOBC}
\begin{figure}[t]
\includegraphics[width=0.27\textwidth]{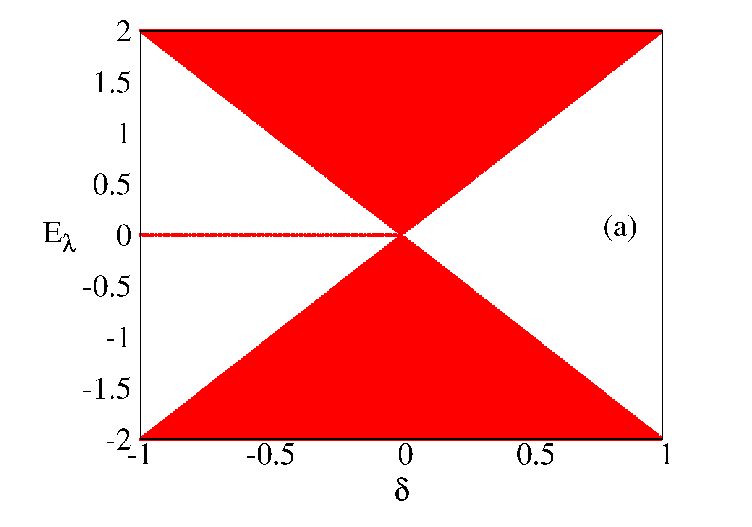}
\includegraphics[width=0.27\textwidth]{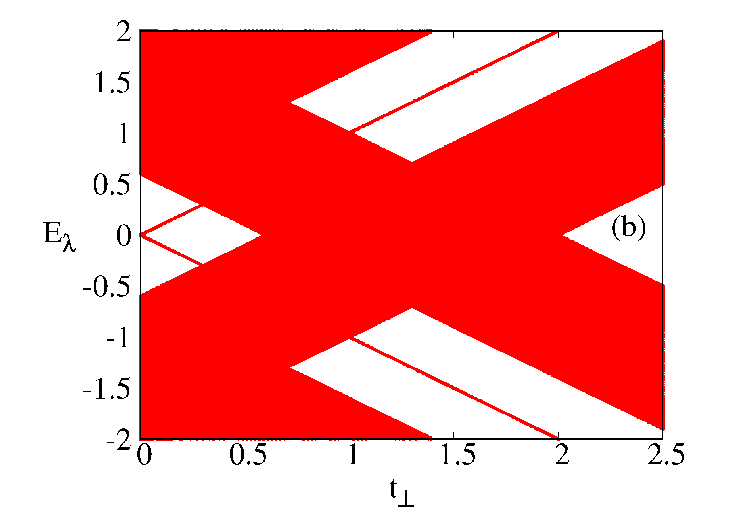}
\includegraphics[width=0.27\textwidth]{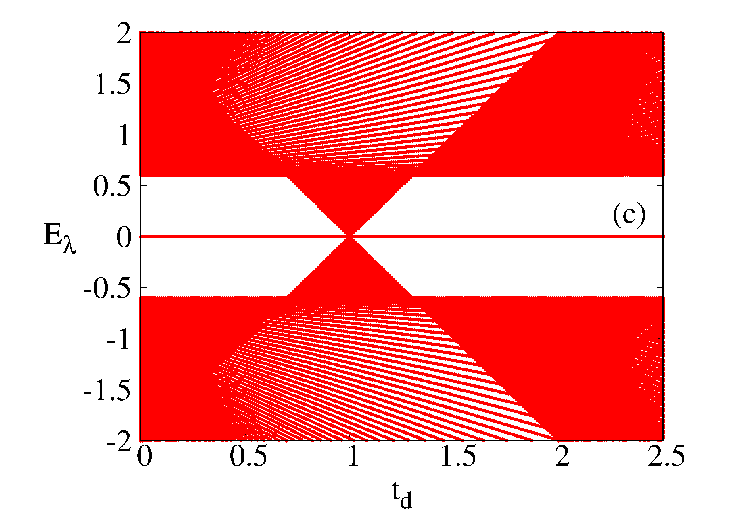}
\caption{\label{fig:EnergySpectrumOBCNoSubstrate} Energy spectrum of single and two SSH wires with OBC and $N_u=200$ (a) as function of $\delta$ for single wire, (b) as function of $t_{\perp}$ for two perpendicularly coupled wires with $\delta=-0.3$ and (c) as function of $t_d$ for two diagonally coupled wires with $\delta=-0.3$.
}
\end{figure}

The nontrivial topological phases in 1D are distinguished by localized edge states with zero energy eigenvalue, for systems with OBC in real space at thermodynamic limit, such that the number of the localized states at each edge is $|W|$. The edge localization is distinguished by the enhancement of spectral weight of the zero energy eigenstates at the edge of the wires, with fast decay while moving away from the edge. Strictly speaking, the localized edge states will have exactly zero energy eigenvalue in the thermodynamic limit. For finite $L_x$, they have energy eigenvalues very close to zero energy for large enough $L_x$. Figure~\ref{fig:EnergySpectrumOBCNoSubstrate}(a) shows the energy eigenvalues $E_{\lambda}$ as function of $\delta$ for a single SSH wire with $L_x=400$ ($\lambda$ is the eigenvalue index). The energy eigenvalues of localized edge states are very close to $E_{\lambda}=0$ for $\delta<0$ and disappear for $\delta>0$, in consistence with the topological phases of the single SSH wire discussed in Sec.~\ref{sec:PhaseDiagramNoSubstrate}.
There are no localized edge states with zero energy eigenvalue for perpendicularly coupled wires with OBC. Figure.~\ref{fig:EnergySpectrumOBCNoSubstrate}(b) displays the energy spectrum for two perpendicularly coupled wires with $\delta=-0.3$ as function of $t_{\perp}$. The edge states of each wire at $t_{\perp}=0$ split for $t_{\perp}\neq0$ into four states, which have the energy eigenvalues $E_{\lambda}=\pm t_{\perp}$. This is due to the perpendicular hopping acting as chemical potential in Eq.~(\ref{eq:HamiltonianPerpBondingAntibonding}). Again, the symmetries of BDI class are only preserved for the combined bonding and antibondig Hamiltonians, therefore, the two negative energy bonding states are the chiral partners of the two positive energy  antibonding states. This confirm the absence of nontrivial topological phase in the perpendicularly coupled wires. These four states disappear for $\delta=0.3$ (not shown).
Figure~\ref{fig:EnergySpectrumOBCNoSubstrate}(c) displays the energy spectrum for diagonally coupled wires with $\delta=-0.3$ as function of $t_d$. It reveals the transition between two nontrivial phases where the critical point is at $t_d=1$. For $t_d<1$, there are two localized states in each edge at zero energy, while for $t_d>1$, there is one localized state in each edge at zero energy. The zero energy localized edge states disappear for $\delta=0.3$ and $t_d<1$, but only two remain for $\delta=0.3$ and $t_d>1$, in consistence with the phase diagram of two diagonally coupled wires (not shown).

\begin{figure}[t]
\includegraphics[width=0.27\textwidth]{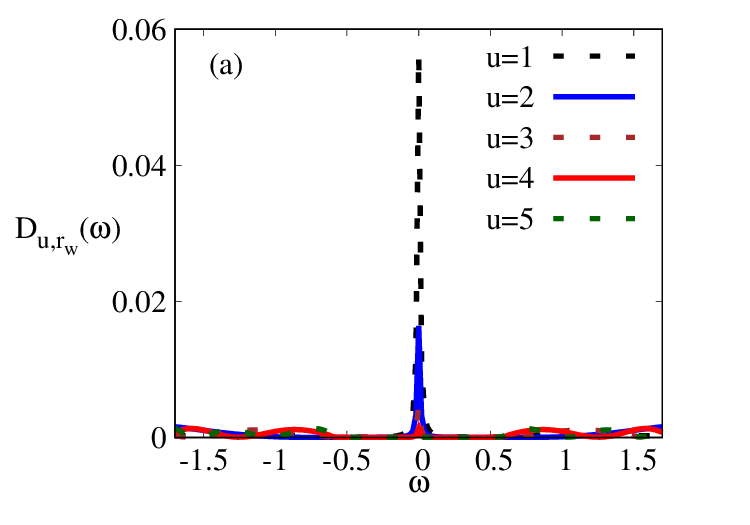}
\includegraphics[width=0.27\textwidth]{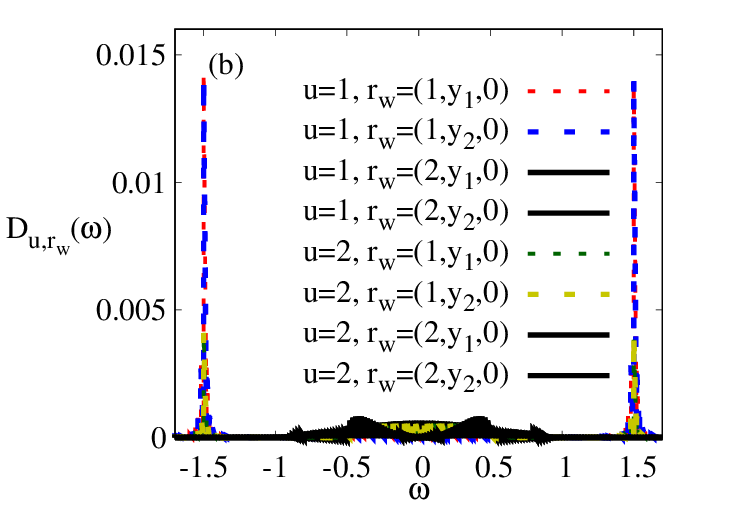}
\includegraphics[width=0.27\textwidth]{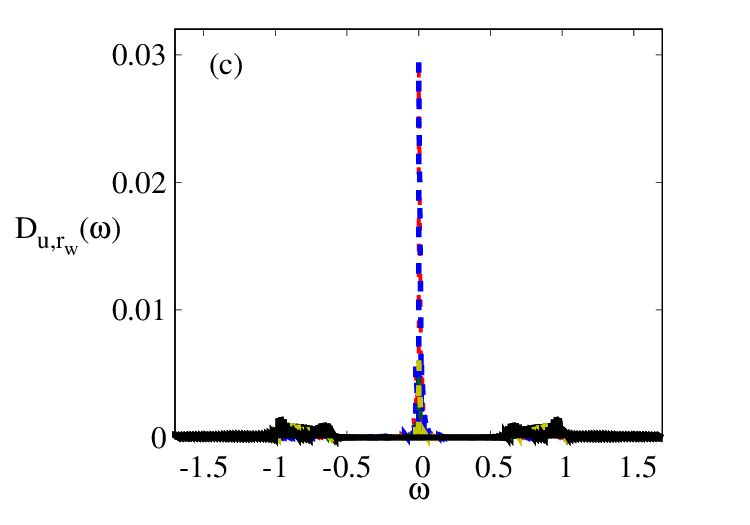}
\caption{\label{fig:LDOSNoSubstrate} LDOS as defined in Eq.~(\ref{eq:LDOS}) at one edge of single and two SSH wires with $\delta=-0.3$ and $N_u=200$. (a) LDOS on sites from the same sublattice in first five unit cells at one edge of single SSH wire (b) LDOS of first two unit cells of two perpendicularly coupled SSH wires with $t_{\perp}=1.5$ (c) LDOS of first two unit cells of two diagonally coupled SSH wires with $t_d=0.5$. (b) and (c) share the same lines key.
}
\end{figure}

The localized edge states can be identified using the local density of states (LDOS) defined as
\begin{equation}\label{eq:LDOS}
 D_{u,r_{\text{w}}}(\omega) = \sum_l |\psi_l(u,r_{\text{w}})|^2 \delta(\omega - E_{\lambda}) ,
\end{equation}
where $| u,r_{\text{w}} \rangle$ are the real space basis representing the wire, and $\psi_l(u,r_{\text{w}})=\langle u,r_{\text{w}} | \phi_l \rangle$, such that $| \phi_l \rangle$ are the energy eigenvectors of $H$. The LDOS corresponds to spectral lines observed in the scanning tunneling spectroscopy measurements of surfaces.
Figure~\ref{fig:LDOSNoSubstrate}(a) displays the LDOS at one edge of single SSH wire with $\delta=-0.3$ and $L_x=400$. The largest spectral weight at zero energy is on the site at the edge of the wire, ie. first site at $u=1$. The spectral weight decays rapidly by going to the following unit cells in the same sublattice and vanishes in the sites of the other sublattice. The other edge displays identical LDOS spectral lines by interchanging the sublattices (not shown).
Figure~\ref{fig:LDOSNoSubstrate}(b) displays the local density of states at one edge of two perpendicularly coupled wires without substrate with $\delta=-0.3$ and $t_{\perp}=1.5$. The localized edge states shift from $\omega=0$ energy to $\omega=\pm1.5$.
Figure~\ref{fig:LDOSNoSubstrate}(c) displays the local density of states at the edges of two diagonally coupled wires with $\delta=-0.3$ and $t_d=0.5$. The spectral weight has a peak inside the gap at $\omega=0$ due to the two localized edge states. The spectral weight remains with the peak inside the gap at $\omega=0$ for $t_d=1.5$ (not shown), but with smaller weight than the $t_d=0.5$ case due to only one localized state at the edge.

\section{SSH wires on substrate}\label{sec:SSHWiresWithSubstrate}
We discuss the effect of hybridizing the former wires to the semiconducting substrate, in such a way that respects the symmetries required for the BDI topological insulators. The single SSH wire can be positioned arbitrary parallel to the $x$-direction, such that each wire site is on top of the adjacent surface site.  The perpendicularly and diagonally coupled wires can be positioned in a similar way along the $x$-direction, but respecting the restrictions imposed by the symmetries of BDI class. We chose the two wires to be the closest adjacent perpendicularly (diagonally) coupled wires that hybridize with laterally nearest neighbor (next nearest neighbor) sites on the substrate, similar to Fig.~\ref{fig:WiresOnSubstrate}.
In the following we compare these three different wire-substrate constructions with the former free standing wires.
\subsection{Band structures and single particle spectral functions}\label{sec:BandsSpectralFunctionsWithSubstrate}
\begin{figure}[t]
\includegraphics[width=0.27\textwidth]{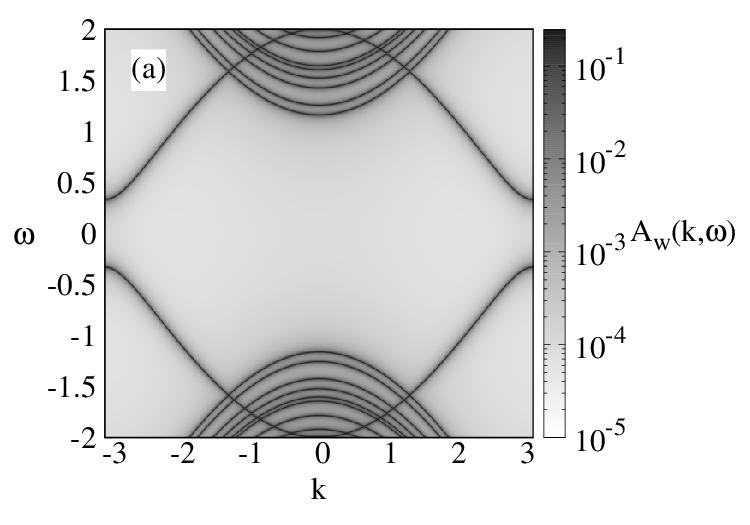}
\includegraphics[width=0.27\textwidth]{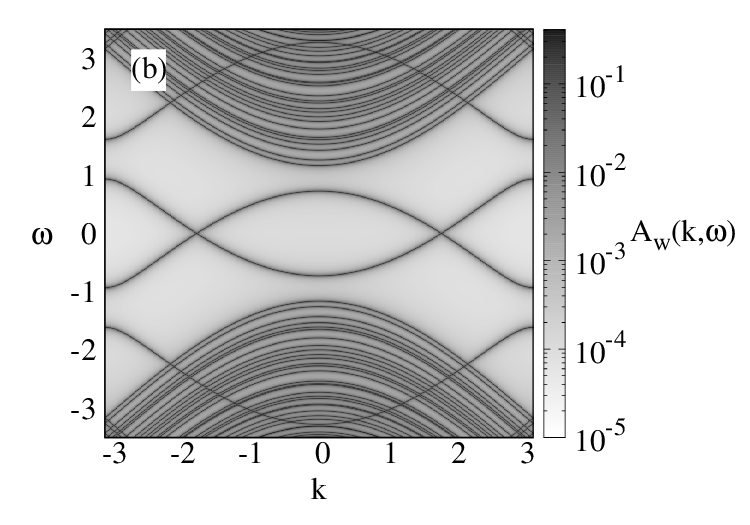}
\includegraphics[width=0.27\textwidth]{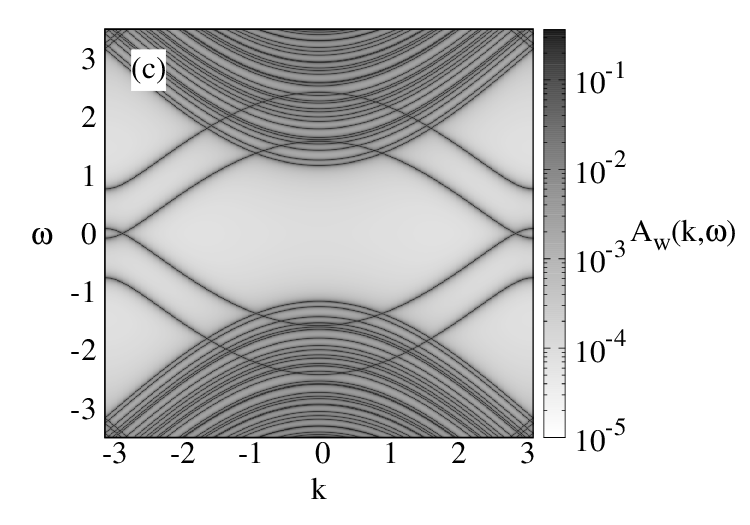}
\includegraphics[width=0.27\textwidth]{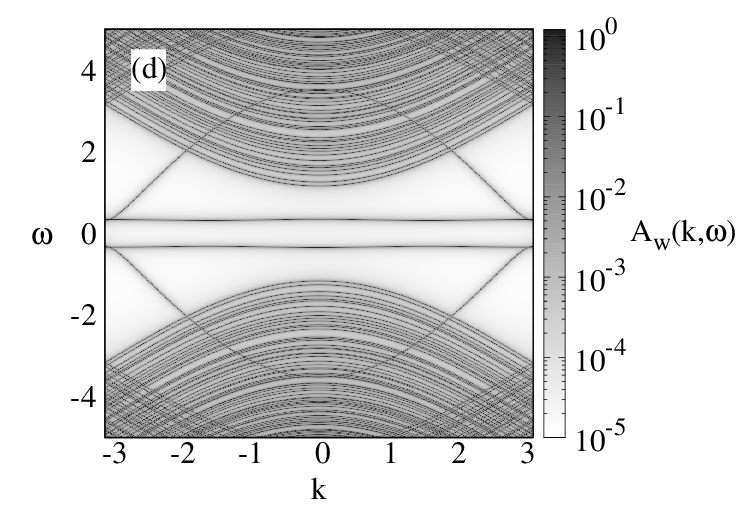}
\includegraphics[width=0.27\textwidth]{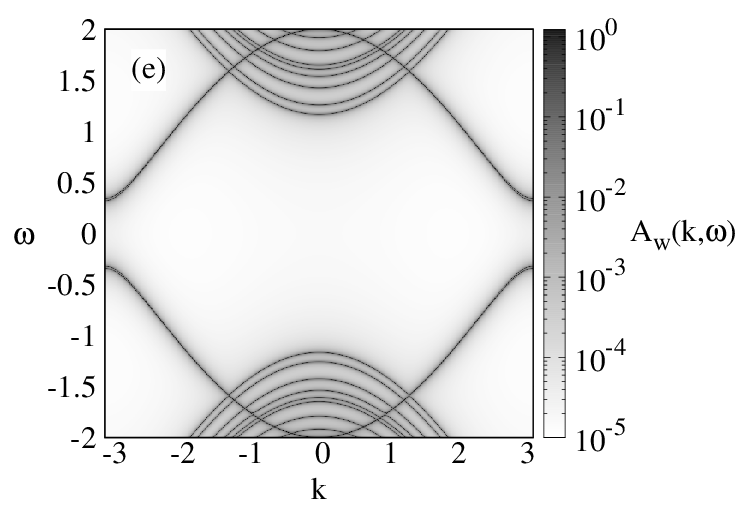}
\caption{\label{fig:SpectFuncWiresWithSubstrate} Single particle spectral function for single and two SSH wires on semiconducting substrate with $N_u=500$, $L_y=16$, $L_z=8$, $\delta=-0.3$ and $t_{\text{ws}}=4$. (a) single wire (b) two perpendicularly coupled wires with $t_{\perp}=1.5$ (c) two perpendicularly coupled wires with $t_{\perp}=0$ (d) two diagonally coupled SSH wires with $t_d=1.5$ (e) two diagonally coupled wires with $t_d=0$.
}
\end{figure}

Figure~\ref{fig:SpectFuncWiresWithSubstrate} shows the single particle spectral function for wire-substrate systems with $L_x=1000$, $L_y=16$, $L_z=8$ and $\delta=-0.3$. In Fig.~\ref{fig:SpectFuncWiresWithSubstrate}(a), the spectral function $A(k,\omega)$ is calculated for a single SSH wire with wire-substrate hybridization $t_{\text{ws}}=4$. We can distinguish two bands with spectral dispersion similar to those of the free standing single SSH wire, but with broadened spectral line and weaker spectral weight due to the hybridization to the substrate. For very weak hybridization, the two bands are very close to those seen in Fig.\ref{fig:SpectralFunctionsNoSubstrate}(a). By increasing the wire-substrate hybridization to $t_{\text{ws}}=4$, we realize the reduction of the band gap, while the top (bottom) of the upper band (lower band) remains without significant change.
The more we increase the wire-substrate hybridization the more these bands approach the band dispersion of the single wire with $\delta=0$. Therefore, we deduce that the wire-substrate hybridization reduces the dimerization effectively towards $\delta^{eff} = 0$. It does not change the basic nature of the intra-wire hoppings.

The two SSH wires on the substrate can be transformed to the bonding and antibonding representations. Then, the wire-substrate hybridization takes the form
\begin{eqnarray}\label{eq:BondingAntibondingWSHybridization}
 H_{\text{ws}} &=& \frac{t_{\text{ws}}}{\sqrt{2}} \sum_{\text{s}=\text{v},\text{c}} \sum_{u}  \left[ \left( c^{\dag}_{u,{\text{s}}, \mathbf{r}}  \left(d^{(+)\phantom{\dag}}_{u,1} + d^{(-)\phantom{\dag}}_{u,1} \right) + \text{H.c.}  \right) \right. \nonumber \\
 &+& \left .\left( c^{\dag}_{u,{\text{s}}, \mathbf{q}}  \left(d^{(+)\phantom{\dag}}_{u,2} + d^{(-)\phantom{\dag}}_{u,2} \right) + \text{H.c.}  \right) \right. \nonumber \\
  &+& \left. \left( c^{\dag}_{u,{\text{s}}, \mathbf{r}^{\prime}}  \left(d^{(+)\phantom{\dag}}_{u,1} - d^{(-)\phantom{\dag}}_{u,1} \right) + \text{H.c.}  \right) \right. \nonumber \\
  &+& \left. \left( c^{\dag}_{u,{\text{s}}, \mathbf{q}^{\prime}}  \left(d^{(+)\phantom{\dag}}_{u,2} - d^{(-)\phantom{\dag}}_{u,2} \right) + \text{H.c.}  \right) \right] , \nonumber \\
\end{eqnarray}
where $\mathbf{r}=(1,1,1)$, $\mathbf{q}=(2,1,1)$, $\mathbf{r}^{\prime}=(1,2,1)$ and $\mathbf{q}^{\prime}=(2,2,1)$. The transformation of the two wires to the bonding and antiboding representation mixes the two sublattices in the substrate hybridization to the perpendicularly coupled wires, but it does not mix them in the hybridization to the diagonally coupled wires.
Figures~\ref{fig:SpectFuncWiresWithSubstrate}(b) and (c) show single particle spectral functions of two SSH wires with $\delta=-0.3$ hybridized with nearest-neighbor sites on the surface. Figure~\ref{fig:SpectFuncWiresWithSubstrate}(c) shows the case of perpendicularly coupled wires with $t_{\perp}=1.5$ and $t_{\text{ws}}=4$. We can distinguish two bonding and two antibondig bands similar to those of free standing perpendicularly coupled wires. For very weak perpendicular hopping, the two bonding (antibondig) bands are very close to those of free standing two perpendicularly coupled wires displayed in Fig.\ref{fig:SpectralFunctionsNoSubstrate}(b), with almost the same energy shift from the bands of a single SSH wire. The absolute value of these energy shifts decreases by increasing the wire-substrate hybridization to $t_{\text{ws}}=4$, which indicates a reduction in the effective perpendicular hopping  $t^{eff}_{\perp}$. Moreover, the gap inside the related bonding (antibonding) bands reduces, which indicates a reduction of the effective dimerization $|\delta^{eff}|$. Figure~\ref{fig:SpectFuncWiresWithSubstrate}(c) shows the case of $t_{\perp}=0$ and $t_{\text{ws}}=4$. At very weak perpendicular hopping, the two band dispersions are very close to those of the free standing single SSH wire seen in Fig.~\ref{fig:SpectralFunctionsNoSubstrate}(a). By increasing the wire-substrate hybridization to $t_{\text{ws}}=4$, we realize a splitting of the two bands into four bands related to the bonding and antibonding bands seen in free standing perpendicularly coupled wires, which indicates the increase of effective wire-wire coupling $t^{eff}_{\perp}$. In fact, similar splitting exists for the case of $t_{\perp}=0$ and finite but small value of $t_{\text{ws}}$, but it is very small to be observed. Therefore, in the absence of direct perpendicular hopping, the wire-substrate hybridization can mediate an effective perpendicular hopping. The gap inside the related bonding (antibonding) bands decreases by increasing $t_{\text{ws}}$, which indicates a reduction of the effective dimerization $\delta^{eff}$. Generally, the increase of wire-substrate hybridization reduces the effective dimerization towards $\delta^{eff}=0$, but renormalizes the effective perpendicular hopping towards $t_{\perp}=t_{\text{s}}$. Again, the wire-substrate does not change the basic nature of the intra-wire hoppings, but it also does not change the basic nature of the perpendicular hopping. However, it can mediate an effective perpendicular hopping in the absence of direct one.

Figure~\ref{fig:SpectFuncWiresWithSubstrate}(d) and (e) show single particle spectral functions of two SSH wires with $\delta=-0.3$, hybridized with next nearest neighbor sites on the surface. Figure~\ref{fig:SpectFuncWiresWithSubstrate}(d) shows the case of $t_d=1.5$ and $t_{\text{ws}}=4$. At very weak diagonal hopping, the wire-related band dispersions are very close to those of free standing diagonally coupled two wires seen in Fig.~\ref{fig:SpectralFunctionsNoSubstrate}(c). By increasing the wire-substrate hybridization to $t_{\text{ws}}=4$, we realize the reduction of the bandwidth of each of the four bands, which indicates the reduction of effective diagonal hopping. Moreover, we realize the reduction of the band gap, which indicates the reduction of effective dimerization. Figure~\ref{fig:SpectFuncWiresWithSubstrate}(e) shows the case of $t_d=0$ and $t_{\text{ws}}=4$.  At very weak diagonal hopping, the two band dispersions are very close to those of the free standing single SSH wire seen in Fig.~\ref{fig:SpectralFunctionsNoSubstrate}(a). By increasing the wire-substrate hybridization to $t_{\text{ws}}=4$, we realize that the two bands remain similar to the single SSH wire bands, but with reduced band gap. This behavior indicates that the wire-substrate hybridization reduces the effective dimerization while keeping vanished diagonal hopping. Therefore, in the absence of direct diagonal hopping, the wire-substrate hybridization does not mediate effective diagonal hopping. The bonding and antibonding bands of free standing diagonally coupled wires match at the bands edges with $k=\pm\pi$. Nevertheless, there is a small energy difference at $k=\pm\pi$ between the bonding and antibondig bands for wires with $t_d=0$, $t_d=1.5$ and wire-substrate hybridization $t_{\text{ws}}=4$, however, hardly distinguishable in the single particle spectral function. This behavior indicates a very small difference in the rate of renormalizing the parameters $t_{(\mp)}$ and $t^{\prime}_{(\mp)}$ in Eq.~(\ref{eq:HamiltonianDiagBondingAntibonding}) between bonding and antibondig bands despite vanishing diagonal hopping.
Moreover, the bonding bands of free standing diagonally coupled wires become completely flat at $t_{(-)}=0$ or $t^{\prime}_{(-)}=0$. However, for wires with $t_d=1.5$ and $t_{\text{ws}}=4$ the bonding bands approach to become flat, but the hybridization to the substrate slightly deforms the bonding bands with very weak dispersion as $\sim \cos(2k)$, also hardly distinguishable in the single particle spectral function. However, this behavior does not change the topological phase, as long as the band gap does not close.
Generally, the wire-substrate hybridization to diagonally coupled wires does not change the basic nature of the intra-wire hoppings, but it also does not change the basic nature of the diagonal hopping, and can not mediate effective diagonal coupling in the absence of a direct one.

\subsection{Phase diagrams}\label{sec:PhaseDiagramWithSubstrate}
\begin{figure}[t]
\includegraphics[width=0.27\textwidth]{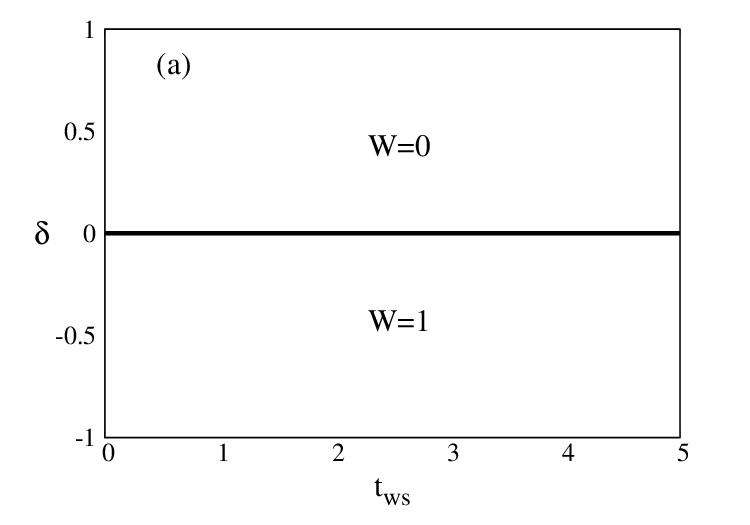}
\includegraphics[width=0.27\textwidth]{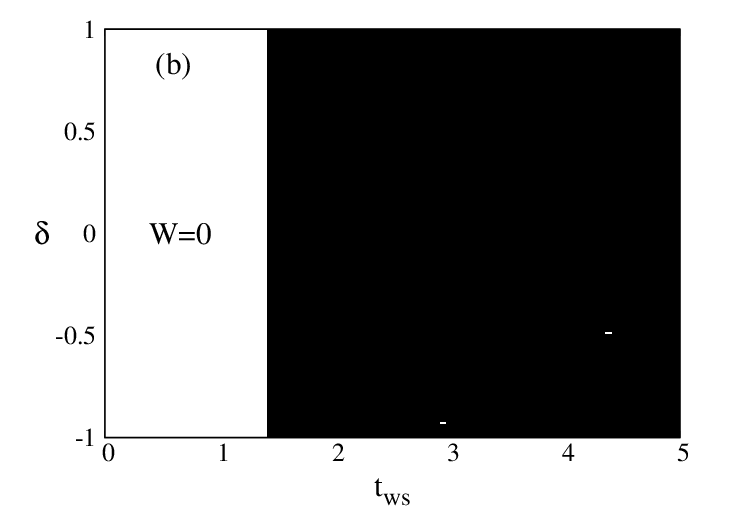}
\includegraphics[width=0.27\textwidth]{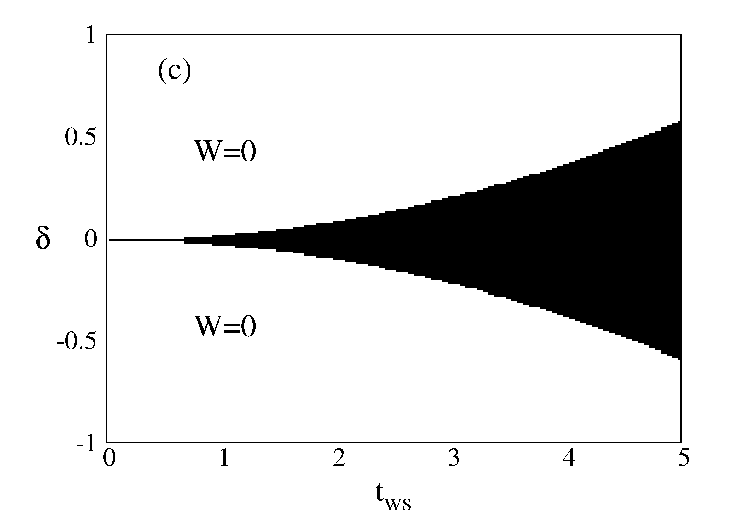}
\includegraphics[width=0.27\textwidth]{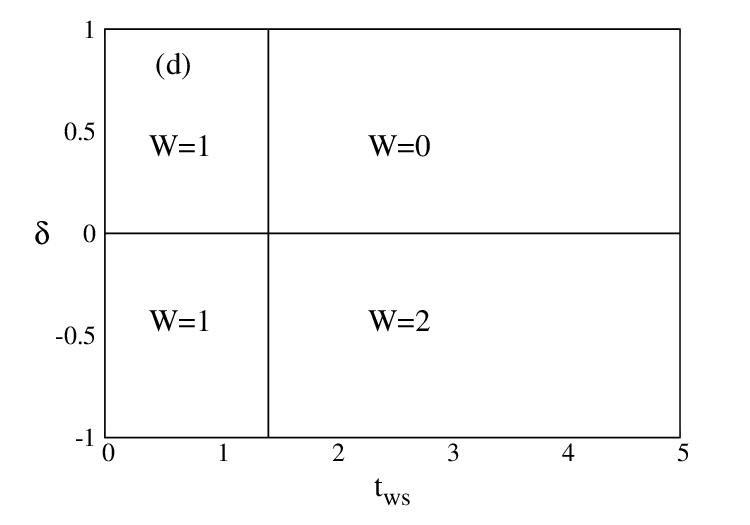}
\includegraphics[width=0.27\textwidth]{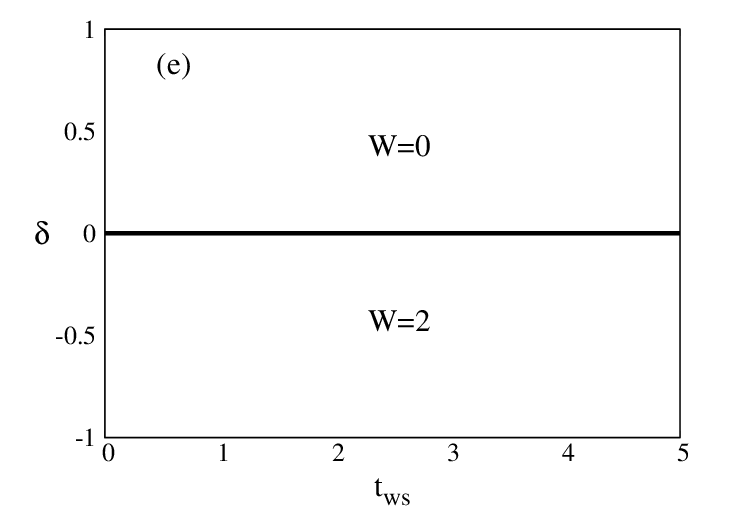}
\caption{\label{fig:PhaseDiagramWithSubstrate} Topological phase diagrams of single and two SSH wires on semiconducting substrate. (a) single wire (b) two perpendicularly coupled wires with $t_{\perp}=2.1$ (c) two perpendicularly coupled wires with $t_{\perp}=0$ (d) two diagonally coupled wires with $t_d=1.1$ (e) two diagonally coupled wires with $t_d=0$.
}
\end{figure}

The wire-substrate hybridization preserves the sign of dimerization, but reduces its absolute value to $|\delta^{eff}|$ until it vanishes at infinitely large hybridization. Thus, the wire-substrate hybridization renormalizes the intra-wire hoppings towards $t_{\text{s}}$, as we observed in the analysis of single particle spectral functions. Therefore, the topological phase of single SSH wire is preserved for finite wire-substrate hybridization. This is demonstrated by calculating the winding number as function of $\delta$ and $t_{\text{ws}}$, which is displayed in Fig.~\ref{fig:PhaseDiagramWithSubstrate}(a). Indeed, the wire-substrate system remains in the topological trivial (nontrivial) phase for $\delta>0$ ($\delta<0$) for finite values of $t_{\text{ws}}$.

The reduction of the effective dimerization and the renormalization of the perpendicular hopping of perpendicularly coupled wires on substrate towards $t_{\text{s}}$ are observed in Figs.~\ref{fig:PhaseDiagramWithSubstrate}(b) and (c). They display the $\delta$-$t_{\text{ws}}$ phase diagram of two SSH wires coupled with $t_{\perp}=2.1$ and $t_{\perp}=0$, respectively. For $t_{\perp}=2.1$ and $t_{\text{ws}}=0$, the system is in a trivial insulating phase with $W=0$, where the gap of the full system depends only on the perpendicular hopping according to the second condition in Eq.~(\ref{eq:BandGapTwoSSHWiresPerp}). By increasing the wire-substrate hybridization, this gap closes, and the system undergoes a phase transition to a gapless phase as expected, due to the renormalization of the wires model parameters towards the model parameter of the substrate, and remains gapless since the uniform perpendicularly coupled wires are in a gapless phase. We can deduce, using Eq.~(\ref{eq:BandGapTwoSSHWiresPerp}), that the critical wire-substrate hybridization does not change by changing the dimerization if the bare perpendicular hopping $t_{\perp}>2t_{\text{s}}$. At $t_{\perp}=0$ and $t_{\text{ws}}=0$, the gap of the full system is the gap due to the dimerization. Increasing the hybridization to substrate mediates effective perpendicular hopping but reduces the effective dimerization. Thus, we can deduce, from the first condition in Eq.~(\ref{eq:BandGapTwoSSHWiresPerp}), that the critical wire-substrate hybridization will increase by increasing the bare dimerization $\delta$. Indeed, we observe, in Fig.~\ref{fig:PhaseDiagramWithSubstrate}(c), a monotonic increase of the gapless region by increasing the wire-substrate hybridization in the $\delta$-$t_{\text{ws}}$ phase diagram.

The reduction of effective dimerization and diagonal hopping in diagonally coupled wires on substrate are observed in Fig.~\ref{fig:PhaseDiagramWithSubstrate}(d). It displays the $\delta$-$t_{\text{ws}}$ phase diagram of two SSH wires coupled with $t_d=1.1$. At $t_{\text{ws}}=0$, the system is in a nontrivial phase with $W=1$. Then, it  undergoes phase transition from topological phase with $W=1$ to topological phase with $W=0$ ($W=2$) for $\delta>0$ ($\delta<0$) by increasing the wire-substrate hybridization. For $t_d=0$, the substrate does not mediate effective inter-wire coupling. Thus the two wires remain in their decoupled wires phase, ie, $W=0$ for $\delta>0$ and $W=2$ for $\delta<0$, which is displayed in Fig.~\ref{fig:PhaseDiagramWithSubstrate}(e).

The wire-substrate hybridization can only drive a topological phase transition between phases that already exist in perpendicularly or diagonally coupled wires, due to renormalization of the bare model parameters. Therefore, the discussion of the phase diagram confirms our previous findings form single particle spectral functions, that the wire-substrate
hybridization does not change the basic nature of the wires model parameters, and it can mediate effective perpendicular hopping but not effective diagonal hopping in the absence of direct wire-wire coupling.

\subsection{Energy spectrum and local density of states with OBC}\label{sec:EnergySpectrumLSOS-WithSubstrateOBC}
\begin{figure}[t]
\includegraphics[width=0.27\textwidth]{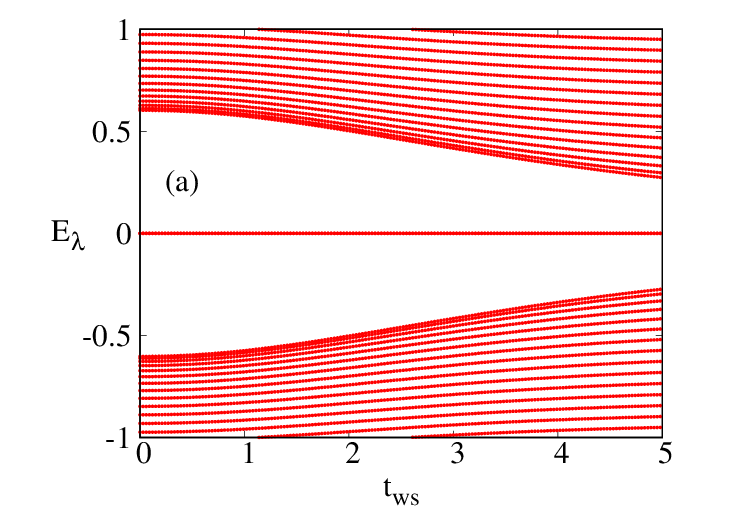}
\includegraphics[width=0.27\textwidth]{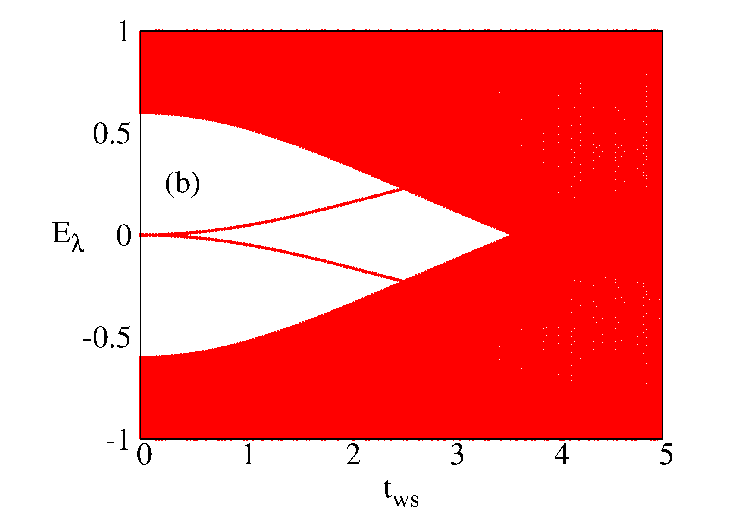}
\includegraphics[width=0.27\textwidth]{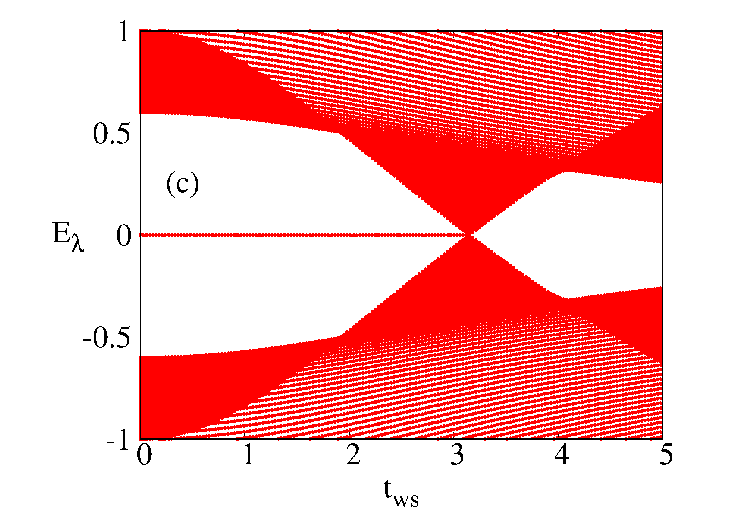}
\caption{\label{fig:EnergySpectrumOBCWithSubstrate} Energy spectrum of single and two SSH wires on a semiconducting substrate with OBC, $N_u=200$, $L_y=4$ and $L_z=4$  as function of $t_{\text{ws}}$ (a) for single wire with $\delta=-0.3$ (b) for two perpendicularly coupled wires with $t_{\perp}=0$ and $\delta=-0.3$ and (c) for two diagonally coupled wires with $t_d=1.5$ and $\delta=0.3$.
}
\end{figure}

\begin{figure}[t]
\includegraphics[width=0.27\textwidth]{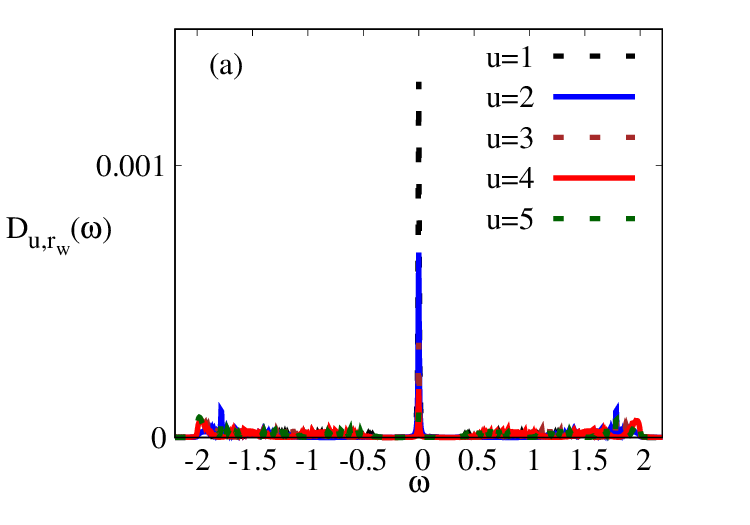}
\includegraphics[width=0.27\textwidth]{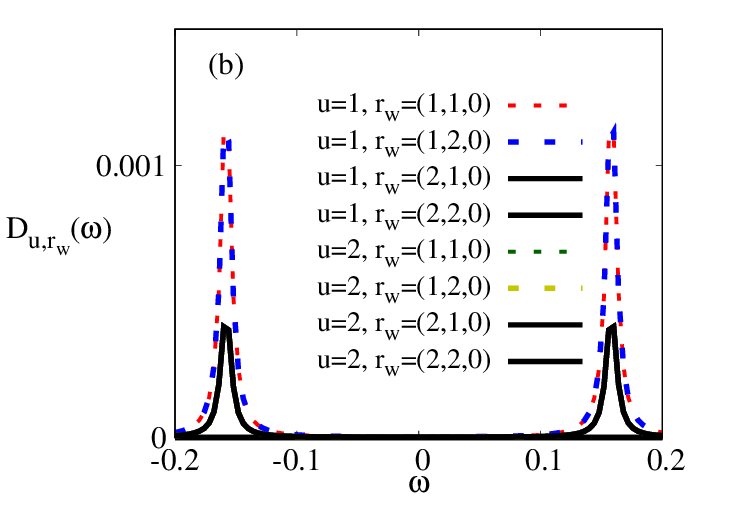}
\includegraphics[width=0.27\textwidth]{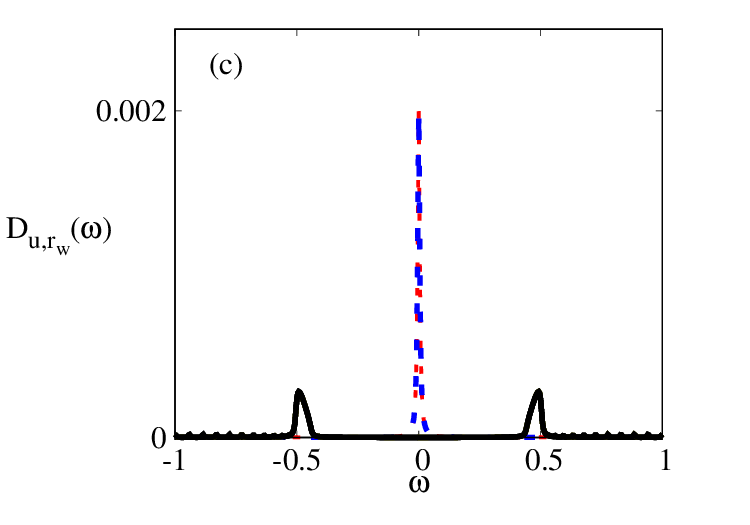}
\caption{\label{fig:LDOSWithSubstrate} Local density of states as defined in Eq.~(\ref{eq:LDOS}) at one edge of single and two SSH wires on semiconducting substrate with OBC, $N_u=200$ $L_y=4$ and $L_z=4$ (a) single wire with $\delta=-0.3$ and $t_{\text{ws}}=4$ (b) two perpendicularly coupled wires with $t_{\perp}=0$, $\delta=-0.3$ and $t_{\text{ws}}=2$ (c) two diagonally coupled wires with $t_d=1.5$ $\delta=0.3$ and $t_{\text{ws}}=2$. (b) and (c) share the same lines key.
}
\end{figure}

The investigation of energy spectrum and LDOS of wire-substrate systems with OBC confirms our findings discussed in the single particle spectral functions and phase diagrams.
The single SSH wire coupled to a substrate with OBC reveals persistence of the localized edge state seen in a free standing wire with $\delta<0$ while increasing the wire-substrate hybridization. This is seen in Fig.~\ref{fig:EnergySpectrumOBCWithSubstrate}(a), which displays the energy spectrum of a single SSH wire, with $\delta=-0.3$ coupled to a substrate, as function of $t_{\text{ws}}$.
The edge states of the single SSH wire are confirmed by calculating the LDOS at one edge of the wire. Figure.~\ref{fig:LDOSWithSubstrate}(a) displays the LDOS on the wire sites, for the SSH wire with $\delta=-0.3$ coupled to the substrate with $t_{\text{ws}}=4$. Most of the spectral weight at zero energy is concentrated at the very last site at the edge, and decreases rapidly by moving away from the edge while keeping in the same sublattice. The spectral weights vanish at the same edge but on sites belonging to the other sublattice.

Two freestanding decoupled wires with $\delta<0$ possess localized edge states at zero energy, two at each edge of each wire. The hybridization of these two wires with the substrate induces effective perpendicular wire-wire hopping, if they are connected to nearest neighbor sites on the surface. This effective perpendicular hopping shifts the energies of the chiral localized edge states away from zero, similar to the bare perpendicular hopping in the free standing perpendicularly coupled wires. The effective perpendicular hopping has a nonlinear relation with the wire-substrate hybridization. Figure.~\ref{fig:EnergySpectrumOBCWithSubstrate}(b) displays the energy spectrum of such two wires with $\delta=-0.3$, where we clearly see the nonlinear shift of chiral states above and below the zero energy, by increasing the wire-substrate hybridization. The energy shift of localized edge states is accompanied with a reduction of the global bulk gap, resulting from the interplay between the reduction of effective dimerization and the increasing of effective perpendicular hopping, similar to the first condition in Eq.~(\ref{eq:BandGapTwoSSHWiresPerp}).
Figure~\ref{fig:LDOSWithSubstrate}(b) displays the LDOS at one edge of this two wire system with $t_{\text{ws}}=2$. The spectral weight at the edge is shifted symmetrically away from the zero energy, due to the effective perpendicular wire-wire hopping.

The number of localized edge states depends on the effective diagonal wire-wire hopping in diagonally coupled wires at fixed dimerization. Figure~\ref{fig:EnergySpectrumOBCWithSubstrate}(c) displays the energy spectrum of diagonally coupled SSH wires on substrate with $\delta = 0.3$ and $t_d=1.5$. At $t_{\text{ws}}=0$, the two wire system possesses one zero energy localized chiral state shared by the two wires at each edge. By increasing the wire-substrate hybridization, we observe the closing of the bulk gap due to the reduction of the effective diagonal hopping. Then, the localized edge states disappear above the critical value of the wire-substrate hybridization, in consistence with the phase diagram in Fig.~\ref{fig:PhaseDiagramWithSubstrate}(d).
Figure~\ref{fig:LDOSWithSubstrate}(c) displays the LDOS at the edge of this two wire system with $t_{\text{ws}}=2$. Most of the spectral weight at zero energy concentrates at the very last two sites at the edge, and decreases rapidly by moving away from the edge while keeping in the same sublattice. The spectral weight vanishes at the same edge but on sites belonging to the other sublattice.

Again, the discussion of the edge states confirm that the wire-substrate
hybridization does not change the basic nature of the wires model parameters, and it can mediate effective perpendicular hopping but not effective diagonal hopping in the absence of direct wire-wire coupling.

\section{Multi wire systems and implications on reconstructions of atomic nanowires on semiconducting surfaces}\label{sec:MultiWireSystems}
\begin{figure}[t]
\includegraphics[width=0.27\textwidth]{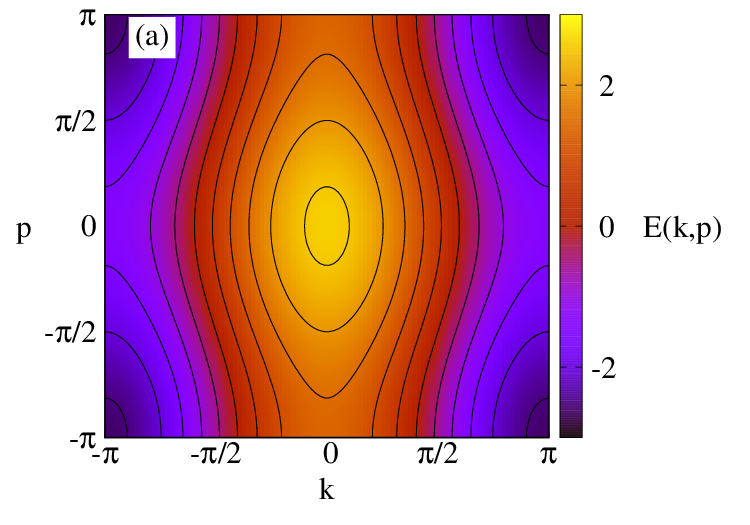}
\includegraphics[width=0.27\textwidth]{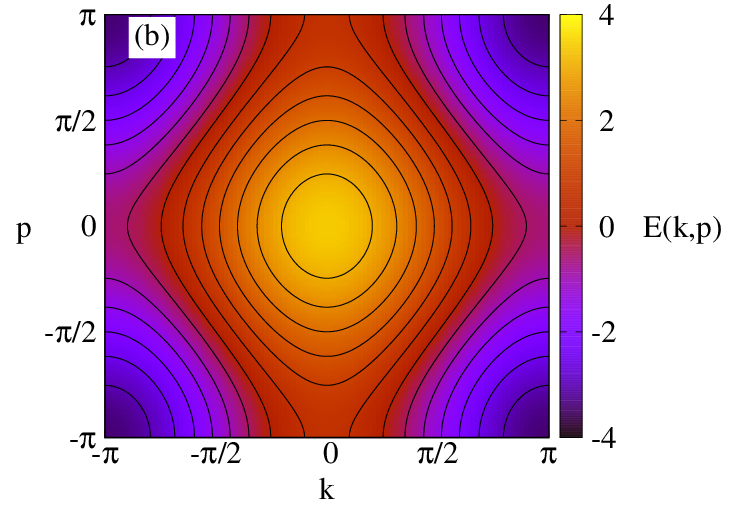}
\includegraphics[width=0.27\textwidth]{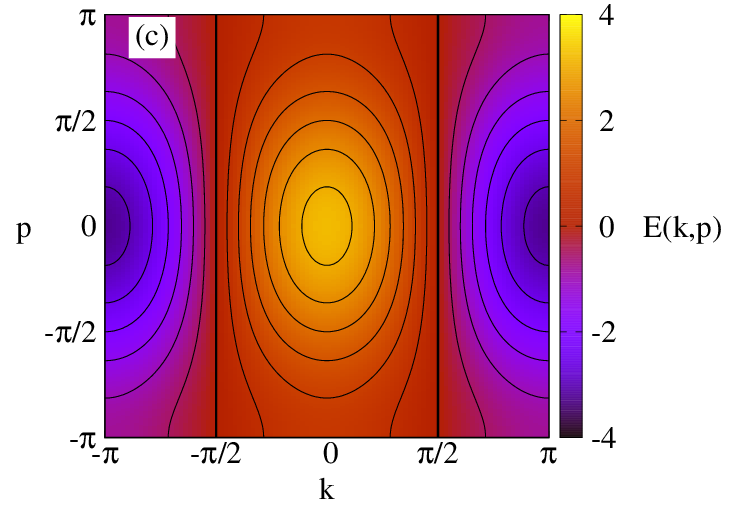}
\includegraphics[width=0.27\textwidth]{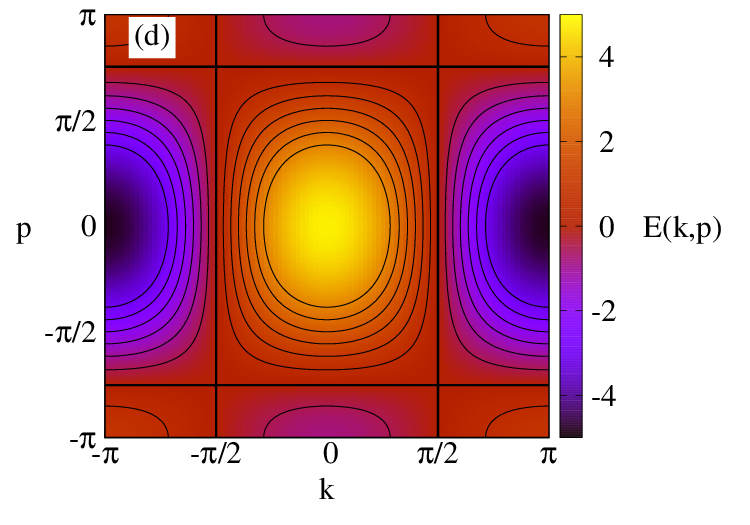}
\caption{\label{fig:Ekp} 
}
(a) and (b) are energy dispersions given by Eq.~(\ref{eq:Dispersion2DPerp}) of two dimensional perpendicularly coupled uniform wires with $t_{\perp}=0.3$ (b) and $t_{\perp}=0.7$, respectively. (c) and (d) are energy dispersion given by Eq.~(\ref{eq:Dispersion2DDiag}) of two dimensional diagonally coupled uniform wires with $t_d=0.3$ (b) and $t_d=0.7$, respectively. The intra wire hopping is $t=1$
\end{figure}

Establishing the effect of the hybridization to the substrate as a renormalization of the wires parameters towards the parameters of the substrate, without changing the basic nature of the model parameters, and finding that the hybridization to the substrate can mediate effective perpendicular hopping but not effective diagonal hopping, in the absence of direct wire-wire coupling, justify neglecting the substrate and discussing only perpendicularly coupled or only diagonally coupled arbitrary number of wires.
So far, we used the SSH model for the wires to establish the impact of the substrate on the wires, using their nature as topological insulators in the BDI class. However, we can restrict our discussion to uniform metallic wires respecting the symmetries of the BDI class.

We consider the wires described by Hamiltonians~(\ref{eq:WiresHamiltonian}), (\ref{eq:perpendicular-hopping}) and (\ref{eq:diagonal-hopping}) with $\delta=0$. Thus, the number of unit cells is $N_u=L_x$, $\text{R}$ in Eq.~(\ref{eq:TransformationMomentumSpace}) reduces to $r_{\text{w}}$ and $t=t^{\prime}=1$. Therefore, by transforming Hamiltonians~(\ref{eq:WiresHamiltonian}), (\ref{eq:perpendicular-hopping}) and (\ref{eq:diagonal-hopping}) to momentum space along the wires direction, we get
\begin{equation}\label{eq:HomogeneousWiresHamiltonianInkspace}
H_{\text{wires}}(k) =  \sum_{\text{w}=1,...,N_{\text{w}}} 2t\cos(k)  c^{\dag}_{k,r_{\text{w}}} c^{\phantom{{\dag}}}_{k,r_{\text{w}}} ,
\end{equation}
\begin{equation}\label{eq:perpendicular-hopping-Inkspace}
 H_{\perp}(k)  =  \sum_{\text{w}=1,...,N_{\text{w}}-1}  t_{\perp} \left( c^{\dag}_{k,r_{\text{w}}} c^{\phantom{{\dag}}}_{k,r_{\text{w}+1}} + \text{H.c.} \right)
\end{equation}
and
\begin{equation}\label{eq:diagonal-hopping-Inkspace}
 H_d(k)  =  \sum_{\text{w}=1,...,N_{\text{w}}-1}  2t_d\cos(k) \left( c^{\dag}_{k,r_{\text{w}}} c^{\phantom{{\dag}}}_{k,r_{\text{w}+1}}+ \text{H.c.} \right) ,
\end{equation}
respectively.
Due to the chiral, time-reversal and particle-hole symmetries, we combine only Eqs.~(\ref{eq:HomogeneousWiresHamiltonianInkspace}) and (\ref{eq:perpendicular-hopping-Inkspace}) (Eqs.~(\ref{eq:HomogeneousWiresHamiltonianInkspace}) and (\ref{eq:diagonal-hopping-Inkspace})) for perpendicularly (diagonally) coupled wires. In both cases, we get the Hamiltonian $H(k)$ in the form of a tridiagonal $N_{\text{w}}\times N_{\text{w}}$ matrix
\begin{equation}
\label{eq:HkMultiwires}
 H(k) =
\begin{bmatrix}
2t\cos(k) & g(k) & 0 & \cdots \\
g(k) & 2t\cos(k) & g(k) & \cdots \\
0 & g(k) & 2t\cos(k) & \ddots \\
\vdots & \vdots & \ddots & \ddots
\end{bmatrix} .
\end{equation}
By diagonalizing Hamiltonian~(\ref{eq:HkMultiwires}), we get the energy bands
\begin{equation}\label{eq:EnergyBandsHomogeneousWires}
 E_{l}(k) = 2t\cos(k) + 2g(k) \cos\left(\frac{l\pi}{N_{\text{w}}+1}\right) ,
\end{equation}
where $l=1,\cdots,N_{\text{w}}$.

For perpendicularly coupled wires, we set $g(k)=t_{\perp}$. Then, the Fermi wavenumber $k_{F_{l}}$ of each energy band $l$ is given by
\begin{equation}\label{eq:BrillouinZoneHomogeneousWiresPerp}
 k_{F_{l}} = \arccos\left(-\frac{t_{\perp}}{t} \cos\left(\frac{l\pi}{N_{\text{w}}+1}\right) \right) .
\end{equation}
Therefore, the number of Fermi wavenumbers is equal to the number of bands. For $N_{\text{w}}\rightarrow \infty$, the Fermi wavenumbers extend in the whole 1D Brillouin zone, ie. $-\pi < k_{F_{l}} \leq \pi$. This is in consistence with 2D perpendicularly coupled wires with PBC in both directions, for which we define the transformation
\begin{equation}\label{eq:TransformationMomentumSpace2D}
 c^{\phantom{\dag}}_{u,r_{\text{w}}} = \frac{1}{\sqrt{N_u N_{\text{w}}}} \sum_{k,p} c^{\phantom{\dag}}_{k,p} \exp \left(-ik u \right) \exp \left(-ip r_{\text{w}} \right) ,
\end{equation}
where the wave numbers $k$ and $p$ are the wave numbers parallel and perpendicular to the wires direction, respectively, defined similar to Eq.~(\ref{eq:wavenumber}). The dispersion relation is given by
\begin{equation}\label{eq:Dispersion2DPerp}
 E(k,p) = 2t\cos(k) + 2t_{\perp}\cos(p) .
\end{equation}
Figure~\ref{fig:Ekp}(a) and (b) show it for $t_{\perp}=0.3$ and $t_{\perp}=0.7$, respectively.
The Fermi wave numbers, parallel ($k_{F}$) and perpendicular ($p_{F}$) to the wires direction, are given by the relation setting $E(k,p) = 0$, where the 1D behaviour is well established for the anisotropic systems with $t_{\perp}\ll t$, at finite temperature with thermal energies well above $\sim t_{\perp}$ but below the energy scale of 2D and 3D orders~\cite{giamarchi04}.

The diagonally coupled wires behave strikingly different. The energy bands are given by
setting $g(k)=2t_d\cos(k)$ in Eq.(~\ref{eq:EnergyBandsHomogeneousWires}).
Thus, we realize that all energy bands have only the two Fermi wavenumbers $k_{F_{l}}=\pm\frac{\pi}{2}$, rendering each of them a strictly 1D effective band. The 2D diagonally coupled wires reveal interesting behavior. The transformation to momentum space in both directions gives rise to the energy dispersion
\begin{equation}\label{eq:Dispersion2DDiag}
 E(k) = E(k,p) = 2t\cos(k) + 4t_d\cos(k)\cos(p) ,
\end{equation}
displayed in Fig.~\ref{fig:Ekp}(c) for $t_d=0.3$ and (d) for $t_d=0.7$.
When $2t_d < t$, the first Brillouin zone has two Fermi lines, extending along the $p$ direction exactly at $k_{F}=\pm \frac{\pi}{2}$. This is a characteristics of strictly 1D bands emerges from 2D system. Nevertheless, the bands are dispersive along the $p$ direction. However, when $2t_d > t$ the Brillouin zone contains four Fermi lines. Two lines extend along the $p$ direction exactly at $k_{F}=\pm \frac{\pi}{2}$ and two lines extend along the $k$ direction exactly at
$p_{F} = \arccos\left( \frac{-t}{2t_d}\right)$ .
This resembles a system of strictly 1D bands along one direction accompanied with other strictly 1D bands along the perpendicular direction, both emerges from 2D system.

Therefore, 2D dispersion alone does not rule out the emergence of strictly 1D behavior as it was suggested in  Au/Ge(001) surface reconstruction~\cite{nak11,par14,jon16}.
The analysis of uniform multi wire systems can facilitate the debate on the experimental results of the Au/Ge(001) surface reconstruction. The crystal structure of the Ge(001) substrate is bipartite. Therefore, it is important to understand how the wires are coupled together and hybridized with the substrate, ie. whether they are perpendicularly-like or diagonally-like coupled wires, or they have other sort of hybridization and wire-wire coupling. However, while it is plausible to assume the time-reversal symmetry in the absence of external magnetic field or magnetic impurities, the electronic band structure of the semiconducting substrate does not generally respect the particle-hole symmetry. From the other side, correlated 1D metals that reveal Luttinger liquid behavior are derived by linearizing the energy bands around Fermi points, rendering free standing 1D correlated conductors particle-hole symmetric at low energies~\cite{giamarchi04,schoenhammer04,gruener2000,solyom10}. Therefore, it is important to investigate the hybridization of metallic wires that respect the particle-hole symmetry to semiconducting substrates that break it. This should be followed/accompanied with abinitio calculation of more realistic models of Au/Ge(001) reconstructions.

From another side, the Bi/InSb(001) reconstruction reveals Luttinger liquid behavior~\cite{oht15}, although it is prepared with large coverage of Bi on the InSb(001) surface. This raises the question on
the strength and nature of the wire-wire coupling. The Bi/InSb(001) is much less investigated as a candidate of possible 1D physics
in comparison to the Au/Ge(001) especially using STM/STS. Indeed, our findings motivate detailed investigations on the exact surface structure to uncover the exact mechanism of emergent Luttinger liquids of 1D correlated metals.

The problem in the previous attempts to discus the existence
of Luttinger liquids in the Au/Ge(001) reconstruction was in trying to make the interpretation
as 1D character "against" 2D character.
However, we think that the existing techniques (STM/STS, ARPES, etc) offer a way to
resolve the debate on the 1D/2D character. The analysis of previous and future experiments have to consider
the possiblilty of emergent exactly 1D character
form 2D dispersive systems, and possible effective diagonal coupling.
The already-seen power-law decay of local density of states in STM and ARPES experiments is a solid evidence for
1D correlated metal~\cite{blu11,dud17}. The nonlocal ARPES  measurements reveal clear 2D dispersion,
but the shape of the Fermi surface is debated. However, the latter has
1D character even in investigations supporting 2D character and ruling out
Luttinger liquid~\cite{nak11,jon16,dud17}.
In order to compare with experiments, we state the need of theoretical investigations of correlated
diagonally coupled chains and explicit calculations of Luttinger liquid
properties.

As example of interacting wires we can consider spinless fermions with nearest neighbor intra-wire interaction similar to that considered in~\cite{abd18,abd21}. Two such interacting wires, coupled with perpendicular hopping, lead to a charge density wave insulator at any finite value of the perpendicular hopping~\cite{giamarchi04}.
However, if the two interacting wires are coupled with diagonal hopping, this leads to two Luttinger liquids with different charge velocities~\cite{giamarchi04}, but up to our knowledge, we did not find analytical investigation on the transition to charge density wave insulator of such system by increasing the interaction. This finding clarifies the results in~\cite{abd21} for two nearest and next-nearest neighbor wires without direct wire-wire coupling. We can safely state that the two nearest neighbor wires are charge density wave insulators due to the substrate mediated perpendicular hopping. The next nearest neighbour wires are decoupled spinless fermion wires since the substrate will not mediate effective diagonal hopping.  However, the substrate still imposes the superpositions of bonding and antibonding fermions in the wires rendering the wire-wire interaction possible among them. Therefore, the next nearest neighbor wires reveal 1D Luttinger liquid phase, but the transition to the charge density wave insulating phase seen in~\cite{abd21} should be clarified in more details. Extending the number of interacting wires with diagonal hopping renders the problem complex. However, the one dimensional behavior in the noninteracting limit can lead to the general form of sliding or crossed sliding Luttinger liquids~\cite{muk01R,muk01,fle18,beg19,geo22,du22}.
Interestingly, a dimensional reduction of a 2D correlated system into a set of correlated one dimensional systems has been found in the presence of both perpendicular and diagonal exchange coupling in spin systems~\cite{sta02,yan12}. The dimensional reduction found there is due to the enhancement of diagonal coupling. This can be seen in the noninteracting limit in which the diagonal hopping acts as an additive term to the intra-chain hopping terms while the perpendicular hopping act as a chemical potential that shifts the bands. Therefore, at vanishing perpendicular term, one ends up with a set of effective 1D systems analogue to what we discussed in diagonally coupled wires. All these finding related to diagonally coupled 1D systems need thorough investigations in the presence of strong correlation.

\section{Conclusion}\label{sec:conclusion}
The hybridization to substrate renormalizes the model parameters, of single and two SSH wires coupled to a simple cubic semiconducting substrate that respects the symmetries of the BDI class, towards the model parameters of the substrate. The substrate can mediate effective perpendicular hopping in the absence of direct perpendicular hopping between the wires, but it can not mediate effective diagonal hopping in the absence of direct diagonal hopping. The hybridization to the substrate does not change the basic nature of the perpendicular or the diagonal hopping.
This behavior justifies neglecting the substrate and considering 2D arrays of perpendicularly or diagonally coupled uniform tight binding wires without dimerization. We established the possibility to realize properties of strictly 1D atomic wires emerging from 2D arrays of diagonally coupled wires, despite strong two dimensional dispersions.
These findings can facilitate the investigations of Au/Ge(001) and Bi/InSb(001) reconstructions, where properties of 1D Luttinger liquid are observed despite strong energy dispersion perpendicular to the direction of the wires or large surface coverage.

\acknowledgments
We thank R. Verresen, H. Frahm, E. Jeckelmann, R. Rausch and K. Khanal for helpful discussions.

\bibliographystyle{unsrt}
\bibliography{bibliography.bib}
\end{document}